\documentclass[journal]{IEEEtran}
\usepackage{color,amsfonts,amsmath,amssymb,booktabs,multirow,psfrag,bm,amsbsy,cite}
\usepackage{enumerate}
\usepackage{subfigure}
\usepackage{changes}
\usepackage{units} 
\ifCLASSINFOpdf
   \usepackage[pdftex]{graphicx}
  \usepackage{epstopdf}
  \DeclareGraphicsExtensions{.pdf,.jpeg,.png}\else  
\fi

\begin{document}
%
\title{Simulation and Measurement Based Vehicle-to-Vehicle Channel Characterization: Accuracy and Constraint Analysis}
\author{Taimoor~Abbas,~\IEEEmembership{Student~Member,~IEEE,}
        J\"{o}rg~Nuckelt,
        Thomas~K\"{u}rner,~\IEEEmembership{Senior~Member,~IEEE,}
				Thomas~Zemen,~\IEEEmembership{Senior~Member,~IEEE,}
        Christoph Mecklenbr\"{a}uker,~\IEEEmembership{Senior~Member,~IEEE,}        
        and~Fredrik~Tufvesson,~\IEEEmembership{Senior~Member,~IEEE}
\thanks{T. Abbas and F. Tufvesson are with the Dept. of Electrical and Information Technology, Lund University, Lund, Sweden. e-mail: ({taimoor.abbas, fredrik.tufvesson}@eit.lth.se).}
\thanks{J. Nuckelt and T. K\"{u}rner are with Institut f\"{u}r Nachrichtentechnik, Technische Universit\"{a}t Braunschweig, Braunschweig, Germany. e-mail: ({nuckelt, kuerner}@ifn.ing.tu-bs.de).}
\thanks{C. Mecklenbr\"{a}uker is  with Institut f\"{u}r Nachrichtentechnik und Hochfrequenztechnik, Technische Universit\"{a}t Wien, Vienna, Austria. e-mail: (christoph.mecklenbraeuker@tuwien.ac.at).}
\thanks{Thomas Zemen is with Forschungszentrum Telekommunikation Wien (FTW), Vienna, Austria. e-mail: (thomas.zemen@ftw.at).}}



\maketitle

\begin{abstract}
In this paper, a deterministic channel model for vehicle-to-vehicle (V2V) communication, is compared against channel measurement data collected during a V2V channel measurement campaign using a channel sounder. Channel metrics such as channel gain, delay and Doppler spreads, eigenvalue decomposition and antenna correlations are derived from the
ray tracing (RT) simulations as well as from the measurement data obtained from two different measurements in an urban four-way intersection scenario. The channel metrics are compared separately for line-of-sight (LOS) and non-LOS (NLOS) situation. Most power contributions arise from the LOS component (if present) as well as from multipaths with single bounce reflections. Measurement and simulation results show a very good agreement in the presence of LOS, as most of the received power is contributed from the LOS component. In NLOS, the difference is large because the ray tracer is unable to capture some of the multi bounced propagation paths that are present in the measurements. Despite the limitations of the ray-based propagation model identified in this work, the model is suitable to characterize the channel properties in a sufficient manner.
\end{abstract}



\IEEEpeerreviewmaketitle
\section{Introduction}

\IEEEPARstart{V}{ehicle-to-Vehicle} (V2V) communication has recently attracted considerable attention from both academia and industry as it facilitates cooperative driving among vehicles for improved safety, collision avoidance, and better traffic efficiency. The radio channel poses one of the main challenges for V2V communication systems design because the V2V channel is highly dynamic. Fast variations in the traffic density, varying vehicle speeds and changing roadside environments, with a height of transmitter (TX) and receiver (RX) antennas relatively close to the ground level makes the V2V channel significantly different from the well studied channels of other technologies such as cellular networks. Thus, a deep understanding of the underlying propagation channels is required. 

In order to characterize V2V channel properties, a number of channel measurements and ray-tracing simulation-based studies have been presented in recent years, e.g., \cite{Wiesbeck2007, molisch09-CommMag, Matolak08, Paier2010a, Nuckelt11c, Mangel011-3, SommerWONS2011, Hosseini2011, Gaugel12}. In addition to that a large body of work on statistical channel modeling with measurement verification is conducted in the past \cite{Chelli2009, Zajic2009a}. Despite these efforts, there is still a need for adequate and reliable, deterministic as well as stochastic channel models allowing a realistic V2V system analysis \cite{molisch09-CommMag, Santos12}. 

The measurement based investigations require a lot of effort and are costly in contrast to the deterministic model based ray-tracing simulations which allow to investigate any desired scenario with less effort and reduced complexity. However, the results obtained from the ray tracing simulations strongly depend on the implemented mathematical models as well as on the accuracy of the data used to describe the environment. Thus, it is necessary to validate the simulations. Most of the measurement and simulation studies in the past have almost exclusively been performed independently except in a few cases \cite{Maurer2004,SommerWONS2011,Joerg2013}. This study is an extension of the work presented in \cite{Joerg2013}, in which the simulation results obtained using a ray-optical model were compared against the results obtained from the DRIVEWAY channel measurements performed in the city of Lund using the RUSK Lund channel sounder \cite{Paier2010a}. The ray-tracing simulator is developed by the researchers at TU Braunschweig especially for vehicular communications at the $5.9$\,GHz band \cite{Nuckelt11c}. The simulations were done for the same urban intersections as where the measurements were performed. In \cite{Joerg2013} the analysis was limited to a single-input single-output (SISO) antenna configuration and a comparison was made only in terms of power delay profile (PDP) and path loss metrics.

\begin{figure*}[t]
     \begin{center}
		
        \subfigure[]{%
            \label{fig:scenario}
            \includegraphics[width=0.33\textwidth]{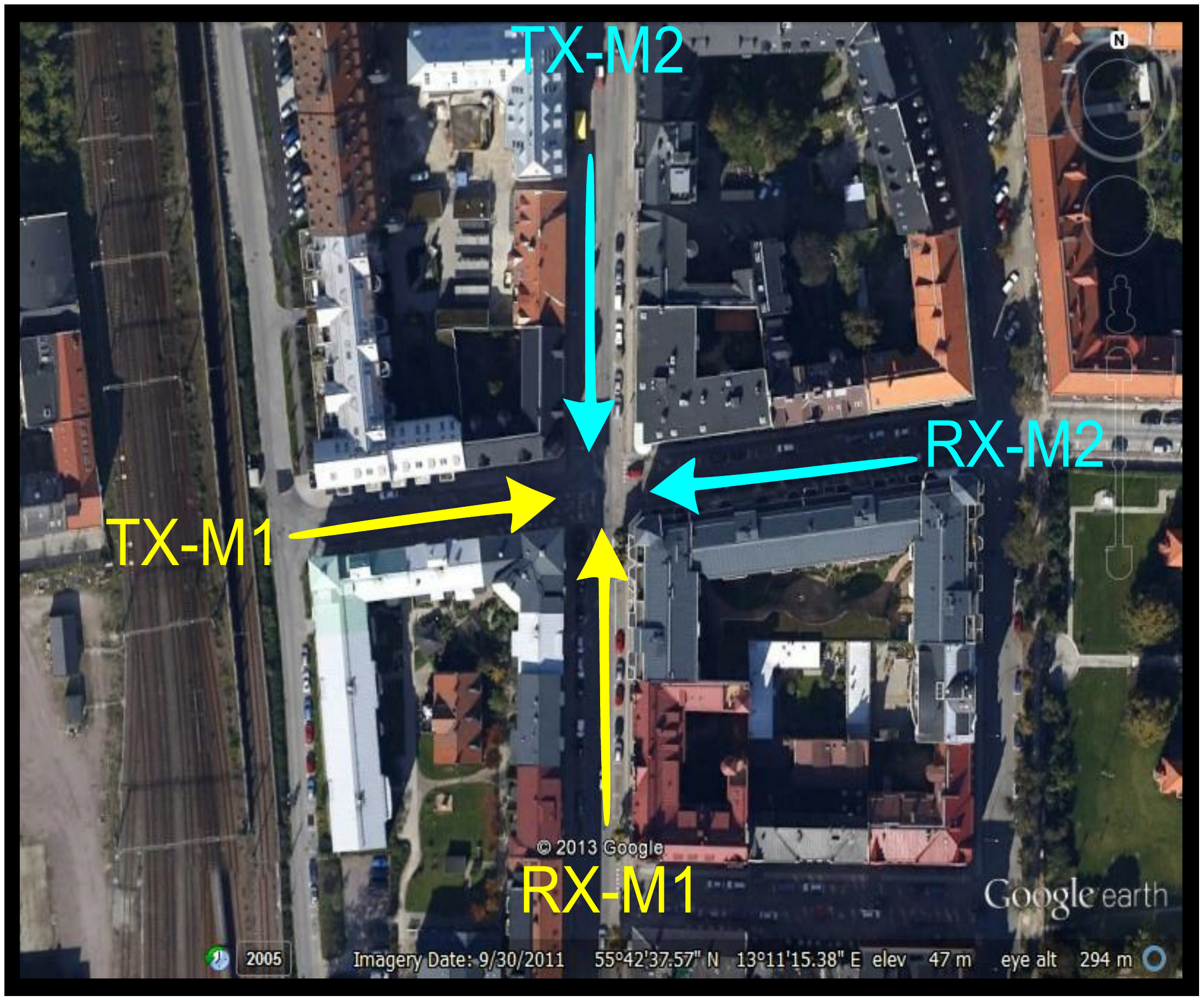}
        }%
        \subfigure[]{%
        		\label{fig:M1}
		\includegraphics[width=0.33\textwidth]{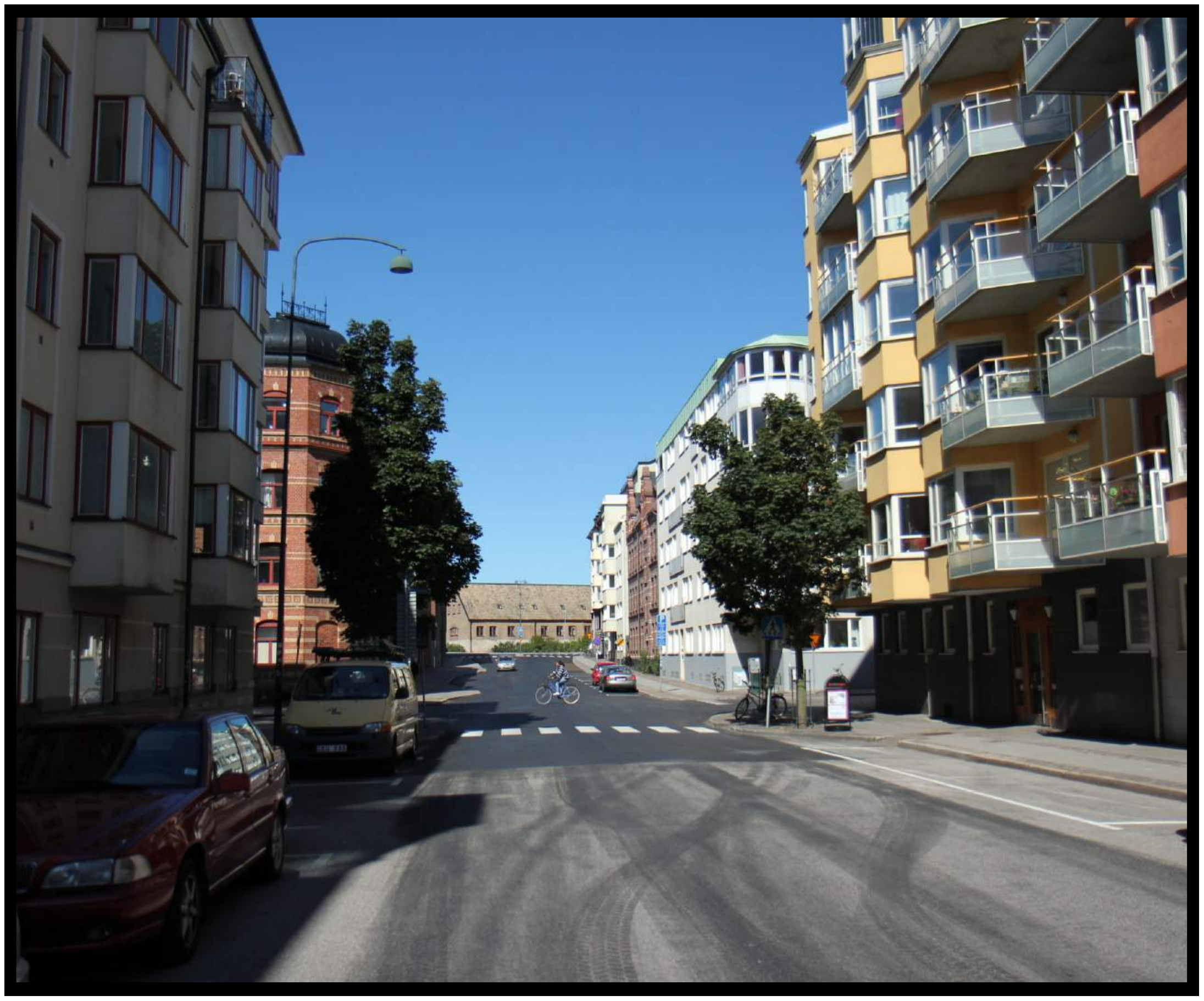}
        }%
         \subfigure[]{%
        		\label{fig:M2}
		\includegraphics[width=0.33\textwidth]{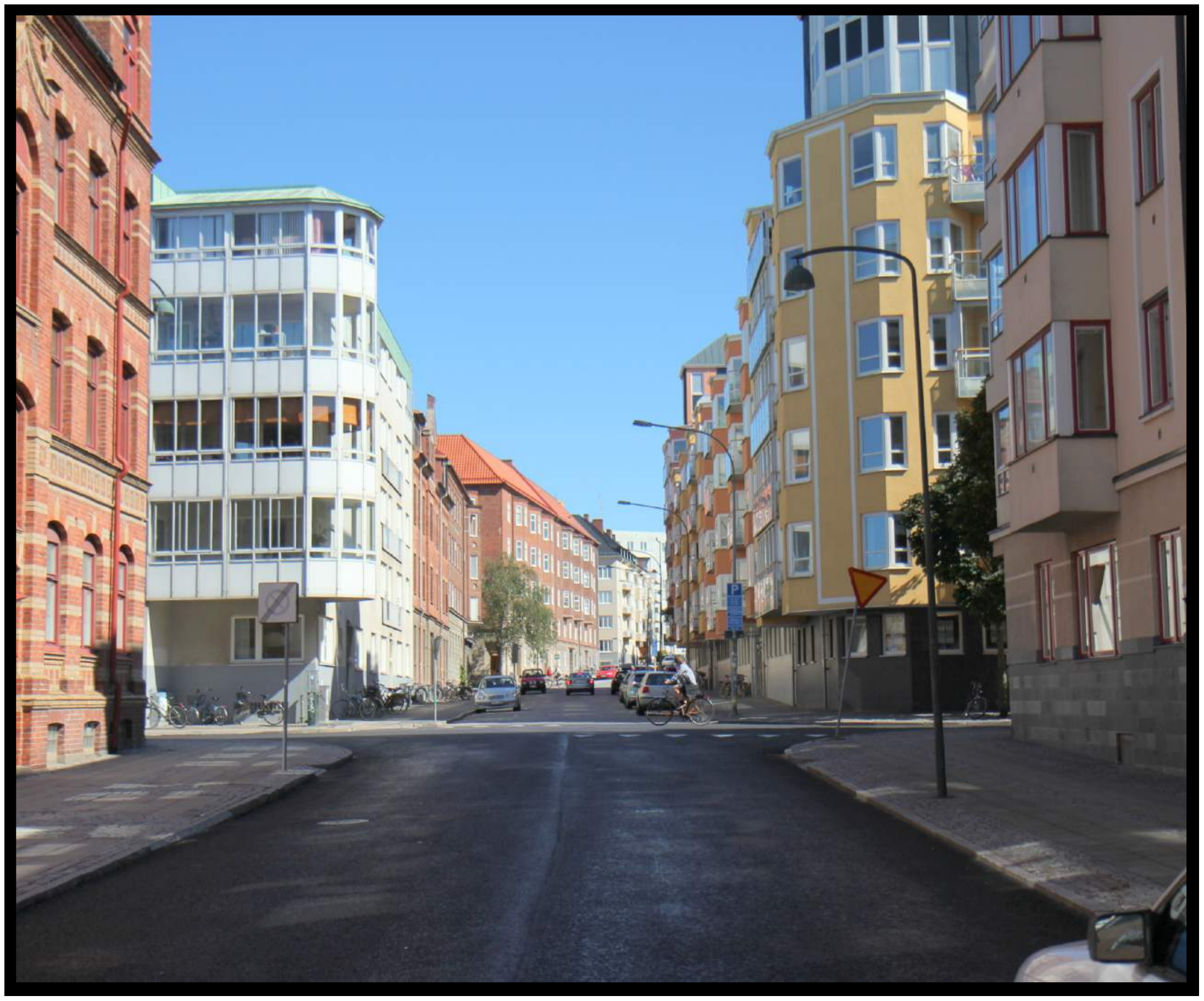}
        }%
        \end{center}
    \caption{%
        (a) Google Earth\textsuperscript{\texttrademark} \cite{GoogleEarth} aerial image of the investigated urban crossroads scenario in the city of Lund with two vehicles moving towards the intersection at a speed of approximately $10$\,m/s (N$55^\circ\,42^\prime\,37^{\prime\prime}$, E$13^\circ\,11^\prime\,15^{\prime\prime}$). (b) and (c) show the perspective from the receiving vehicle $RX-M1$ and $TX-M1$, respectively.
     }%
   \label{fig:Scenarios}
\end{figure*}

The importance of an urban street intersection scenario is that the line-of-sight (LOS) path is often obstructed by surrounding buildings which strongly limits the wave propagation and affects the link reliability in a crucial manner. In such a scenario the communication is highly dependent on the availability of reflected multipath components (MPCs) which in turn depend on the location of the scatterers, street width, distance to the intersection, and side distance of the transmitting node to surrounding buildings. Some path loss models have been derived to characterize such scenarios \cite{Schack11, Mangel011-3} and validated using channel measurement data \cite{Abbas2013ITST}. Both models assume an intersection with perpendicular side roads and cannot directly be applied to intersections with an irregular geometry. Hence, urban intersections constitute an interesting scenario to investigate where a flexible semi-deterministic channel model for V2V communications is still required.\par
\emph{The main contribution of this work} is a detailed evaluation of V2V channel parameters by comparing the results obtained from the ray-tracing simulations and measurements for a $4 \times 4$ multiple-input multiple-output (MIMO) configuration in a four-way-urban street intersection scenario is presented. This paper is a sustained continuation of the work presented in \cite{Joerg2013}, where channel metrics like PDP, channel gain and delay spreads have been analyzed for one V2V scenario. In addition to that, the following channel metrics related to the time-variant nature of the channel and to multi-antenna system are derived in this paper:
\begin{itemize}
\item Doppler spreads,
\item eigenvalue distribution, and
\item MIMO diversity
\end{itemize}
for two different TX and RX configurations are compared against each other. The measurement data is analyzed separately for both LOS and NLOS situations, where mean and standard deviation of the error is provided individually for each of the channel metrics. With the help of these channel metrics the accuracy of the ray-tracing tool is analyzed and possible flaws or limitations of the underlying channel models are identified.

The remainder of the paper is organized as follows: Section \ref{sec:scenario} describes the measured urban intersection scenario as well as the different configurations of the TX and RX vehicles used during the measurements. Section \ref{sec:measurement} describes the RUSK-Lund channel sounder and the measurement setup. In section \ref{sec:simulation} the ray-optical channel model used for the simulations is described. The channel measurement data and the simulation data are compared and analyzed in section \ref{sec:analysis}. Finally, in section \ref{sec:conclusion} the discussion is summarized and conclusions are presented.

\section{Urban Intersection Scenario}
\label{sec:scenario}

For the comparison of simulation results against measurement data, we have chosen an urban four-way intersection (N$55^\circ\,42^\prime\,37^{\prime\prime}$, E$13^\circ\,11^\prime\,15^{\prime\prime}$, see Fig. \ref{fig:scenario}) in the city of Lund, Sweden.


The scenario is exactly the same as the narrow urban scenario described in \cite{karedal10-05}. Two measurements have been selected for the analysis: $M1$) when the TX and RX cars are driving from the streets $TX-M1$ and $RX-M1$, and $M2$) when the TX and RX cars are driving from the streets $TX-M2$ and $RX-M2$, respectively, towards the intersection at a speed of approximately $10$\,m/s. The line-of-sight (LOS) component is obstructed by four-story buildings arranged along each leg of the intersection. On the walls of the buildings there were doors and windows and balconies with metallic frames. The canyon of the street is quite narrow and ranges from $14-17$\,m. The four side roads are not perfectly perpendicular. During the measurements there were parked vehicles along both sides of the streets. Furthermore, there were some traffic signs and lamp posts in the close environment of the intersection. The photos shown in Fig. \ref{fig:M1} and Fig. \ref{fig:M2} show a view from the RX vehicle of the the street $RX-M1$ and the TX vehicle of the street $TX-M1$, respectively. It is worth mentioning that in $M1$ a bus is driving in front of the RX and turns left from $RX-M1$ street to $TX-M1$ street in Western direction at the beginning of the scenario. 

\section{Channel Measurement Setup}
\label{sec:measurement}

The RUSK Lund channel sounder, which performs MIMO measurements based on the switched array principle, was used to record the complex time-varying channel transfer function $H(f,t)\in\mathbb{C}^{M_R\times M_T}$, where $M_T$ and $M_R$ denote the number of transmit and receive antennas, respectively. The corresponding channel impulse response (CIR), $h\left( {\tau ,t} \right)$, is derived from $H(f,t)$ by applying a Hann window to suppress side lobes and then an inverse Fourier transform is performed. For each measurement the sounder sampled the channel for $10$\,s using a time increment of $\Delta t=307.2\,\mu s$ over a bandwidth of $240$\,MHz at $5.6$\,GHz carrier frequency. Two regular hatchback cars with a height of $1.73$\,m and each equipped with a four-element antenna array were used to perform V2V measurements. Each antenna element in the array had a somewhat directional beam pattern pointing; 1) left, 2) back, 3) front, and 4) right, respectively These antenna arrays, integrated into the existing radomes ('shark fins') on the car roof, were specifically designed for V2V communications \cite{thiel10-04}. The interested reader is referred to \cite{Paier2010a}, where a detailed description of the measurement setup can be found. 

\section{3D Ray-optical Channel Model}
\label{sec:simulation}



The radio channel of the V2V scenario described above is characterized by utilizing a deterministic propagation model based on the ray-tracing principle \cite{McKown91}. The ray-based model has been specifically designed for the analysis of V2V communication scenarios \cite{Nuckelt11c, Schack13}. It includes a full 3D representation of the wave propagation and comprises the full-polarimetric antenna gain patterns of the TX and the RX, respectively. Several radio propagation mechanisms are taken into account to properly model the multipath nature of V2V channels. First of all, the LOS path between TX and RX -- if not obstructed by obstacles like buildings -- is calculated based on the distance between both nodes and the free-space path loss formula. Scattering caused by surrounding objects is modeled in two different ways. On the one hand, specular reflections, i.e., the angles of the incident ray and the reflected ray are the same, are calculated by solving the well-known Fresnel equations. The corresponding reflection points are obtained by applying the image method \cite{McKown91}. Note that multi bounce interactions can be theoretically considered up to the $n$-th order in the case of specular reflections. However, for reasons computational complexity the order of specular reflections is practically limited to three or four depending on the complexity of the investigated scenario. On the other hand, scattering in terms of non-specular reflections, i.e., diffuse scattering, is simulated by applying the Lambert's emission law \cite{Degli-Esposti01}. Precisely, the surface of an object that can be seen by both TX and RX is segmented into small tiles and, afterwards, for each tile a single scattering process is solved according to the Lambert's emission law in order to obtain the corresponding scattering coefficients. It is worth mentioning that the current implementation of diffuse scattering accounts for single bounce interactions only. As it is shown later, this is a major limitation of the model when characterization the V2V channel in urban street canyons. Furthermore, the current propagation model does not include diffraction. From the author's point of view it seems reasonable to neglect the contributions caused from diffracted waves at a frequency of 5.9\,GHz (cf. \cite{karedal10-05}).

In order to characterize the V2V radio channel using the ray-based model, the investigated scenario has to be described in detail, including all buildings and obstacles that mainly interact with the transmitted signal and, therefore, affect the wave propagation. For the analysis presented in this manuscript, building data for the design of a virtual scenario is obtained from OpenStreetMap (OSM).\footnote{See www.openstreetmap.org} In \cite{NuckeltEUCAP2013}, a guideline and a proof-of-concept how to make use of OSM building data for 3D ray-tracing simulations is presented. The current development status of the database provides a detailed description of the buildings in metropolitan areas. Note that OSM usually provides only rare information about the individual height of buildings. However, the accurate building height is not essential in this peer-to-peer scenario as most of the wave propagation takes place at street level. For this reason, we have set the height of all building for this analysis to $15$\,m that is close to real height of the four- to five-story buildings that are placed along the roadside. By analyzing videos captured during the measurements and on-site inspections of the intersection, we further identify relevant obstacles like traffic signs, lamp posts or parked cars along the roadside and add these objects in a simplified manner to the virtual scenario.

\begin{figure}[b]
\centering
\includegraphics[width=\columnwidth]{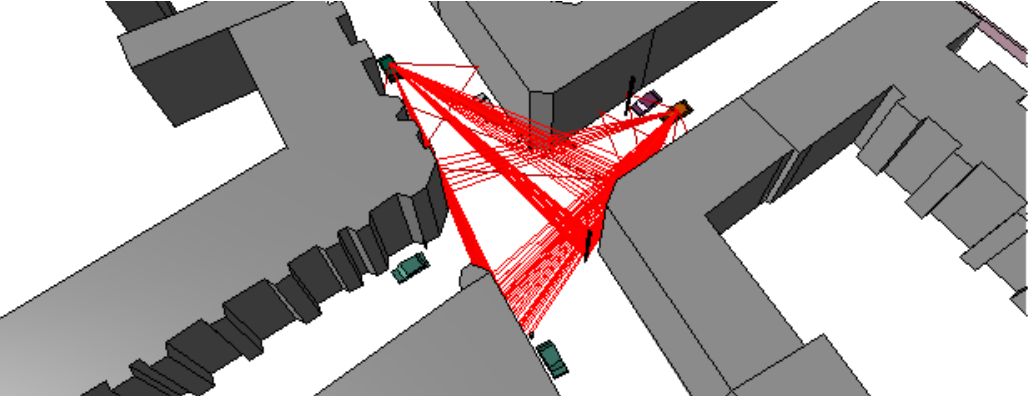}
\caption[]{Simulation of the underlying urban intersection scenario by means of ray tracing. The data of the environment includes buildings, traffic signs, lamp posts as well as parked cars along the roadside.}
\label{fig_raytracing}
\vspace{-0.5cm}
\end{figure}

The positions of the moving TX and RX are reproduced using the GPS coordinates logged during the measurement runs. A time resolution of $10$\,ms has been chosen to initially sample the virtual scenario. For each snapshot the ray-based model determines the propagation paths (rays) including its electro-magnetic properties. Depending on the number of rays, the computation time (accounting for multi bounce specular reflections of order three) for one snapshot is in the range from a few milliseconds up to ten seconds using a standard desktop personal computer. In order to further increase the time resolution of the sampled radio channel, a post-processing routine \cite{Schack13} is applied afterwards. By exploiting the knowledge about the propagation paths determined by the ray-based model it becomes possible to derive the ray information between two adjacent snapshots without using the time-consuming ray tracer. In this way, the time resolution of the sampled scenario is increased to $\unit[100]{\mu s}$ enabling an adequate modeling of the Doppler effect. The intermediate result is a set of rays for each snapshot of the V2V scenario as exemplarily illustrated in Fig.~\ref{fig_raytracing}.

After a coherent superposition of all determined rays, the final output of the ray-based model is the time-variant CIR $h(\tau,t)\in\mathbb{C}^{M_R\times M_T}$, which completely characterizes the frequency-selective channel for each TX/RX link. We can express the CIR as
\begin{align}
 h\left( {\tau ,t} \right) &= \sum\limits_{k = 1}^{N(t)} {a_k (t) \cdot e^{j(2\pi f\tau _k (t) + \varphi _k (t))}  \cdot \delta (\tau  - \tau _k)}  \\ 
  &= \sum\limits_{k = 1}^{N(t)} {\tilde{a} _k (t) \cdot \delta (\tau  - \tau _k)},
  \label{CIR}
\end{align}
where the $k$-th multipath component is described by the amplitude $a_k \left( {t} \right)$, the delay $\tau _k (t)$ and the phase shift $\varphi _k (t)$ at time $t$. $N(t)$ and $f$ denote the number of multipath components for each time instance and the carrier frequency of the system, respectively. Based on the predicted CIR further metrics like the PDP, the channel gain or the RMS delay spread can be derived and compared with measurement-based data.

\section{Analysis}
\label{sec:analysis}

The main objective of this paper is to perform a validation of the ray-tracing simulation results for the V2X channel by means of a comparison with measured channel data. In the simulations, we have restricted the order of specular reflections to second order in order to keep the complexity low. Moreover, non-specular reflection of order higher than one and diffuse scattering are not included either. Given these restrictions and limitations, the goal is to find out how well the channel properties can be described using ray tracing simulations under those conditions. Signal reception is significantly different in LOS and NLOS situations, we thus characterize the channel metrics separately for LOS and NLOS situations. 

\subsection{Power Delay Profile and Doppler Spectral Density}

The time-variant APDP is calculated by using the time-variant $4\times4$ MIMO channel transfer function $H(t,f)$ obtained from the measurement data as well as from the simulations as follows,
\begin{equation}
\label{eq:APDP}
P_\tau(t_k,\tau)=\frac{1}{N_{avg}}\sum_{n=0}^{N_{avg}-1}|h(t_k+n\Delta t,\tau)|^2,
\end{equation} 
where the $P_\tau(t_k,\tau) \in \mathbb{R}^{M_R\times M_T}$ is averaged over a window of $N_{avg}$ time snap-shots such that $N_{avg}\Delta t=57$\,ms corresponding to a TX/RX movement of about $10$ wavelengths at a speed of approximately $10$\,m/s. Furthermore, the processing includes noise reduction of the measurement data as described in \cite{karedal10-05}. 

\begin{figure*}
     \begin{center}
        \subfigure[Measured]{%
            \label{fig:PDP_Meas_17}
            \includegraphics[width=0.48\textwidth]{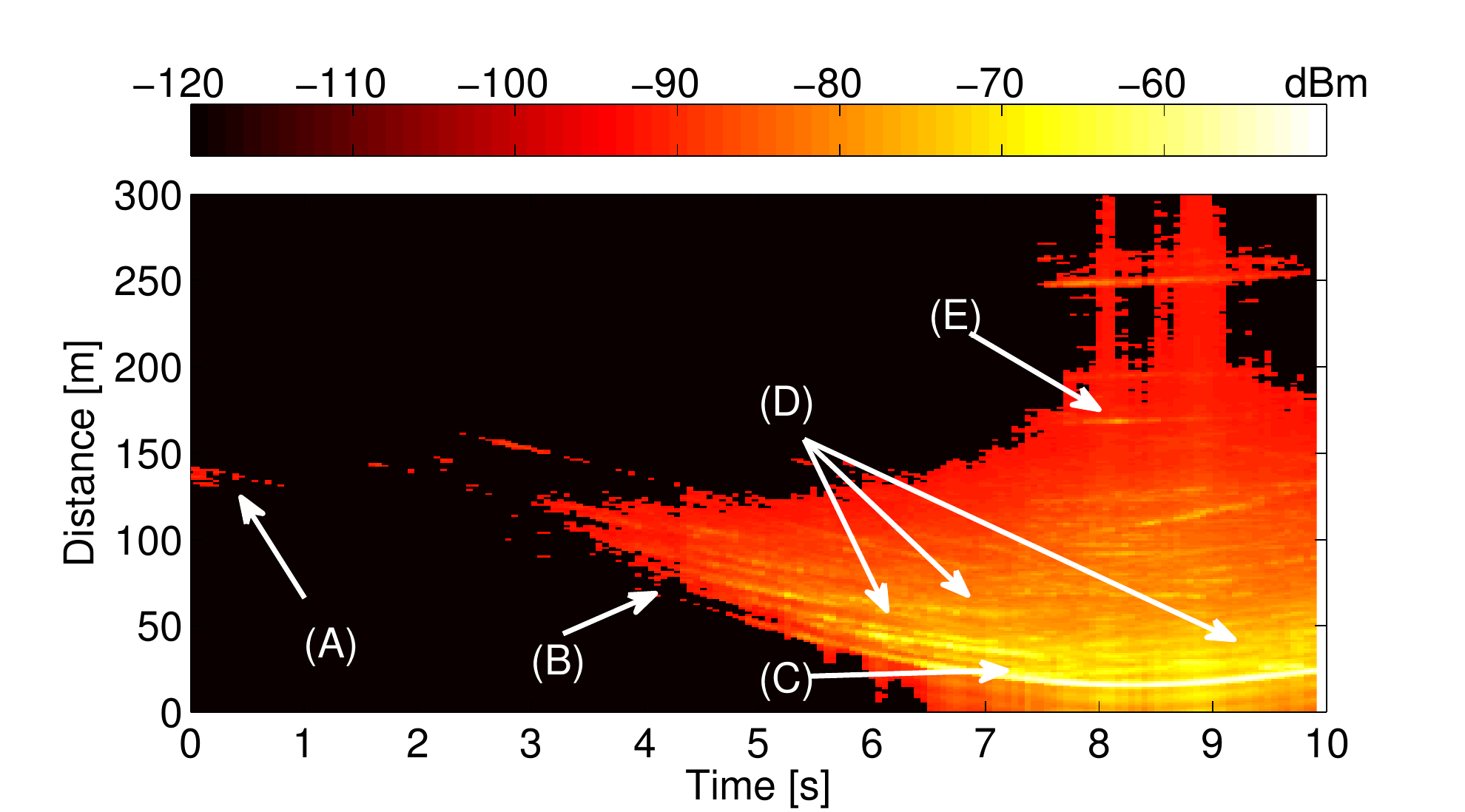}
        }%
		\subfigure[Measured]{%
            \label{fig:PDP_Meas_19}
            \includegraphics[width=0.48\textwidth]{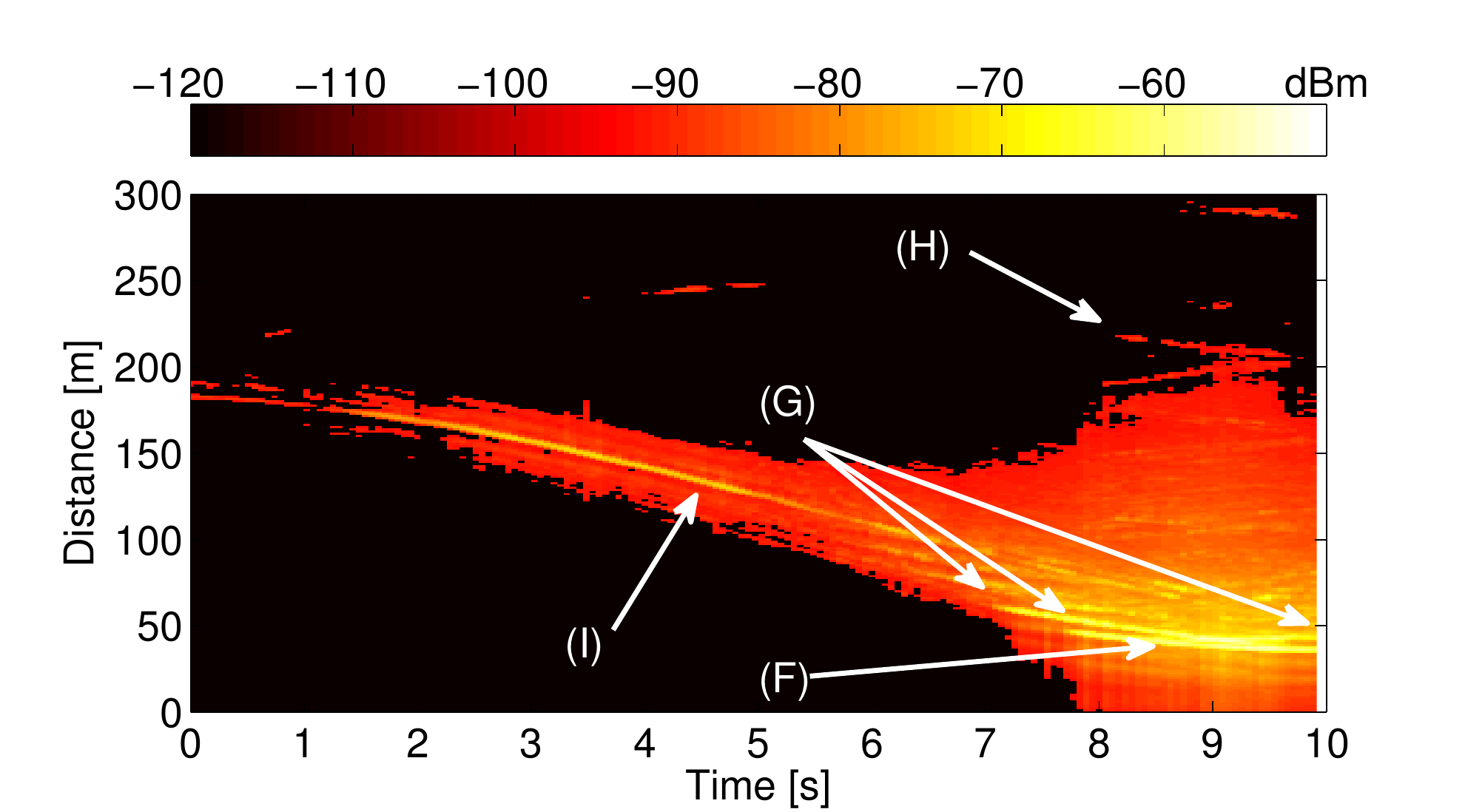}
        }%
        \\
        \subfigure[Simulated]{%
            \label{fig:PDP_RT_17}
            \includegraphics[width=0.48\textwidth]{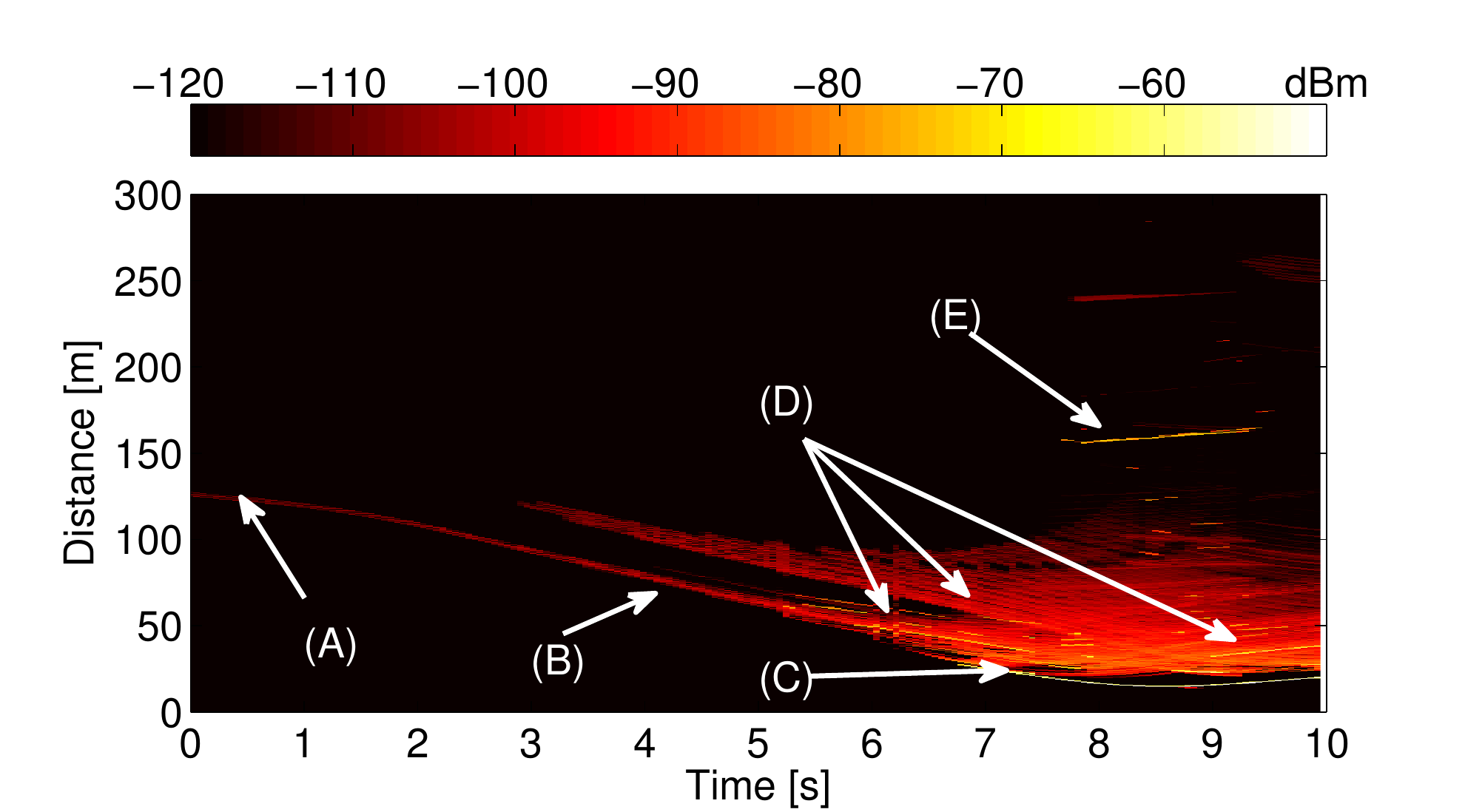}
        }%
		\subfigure[Simulated]{%
            \label{fig:PDP_RT_19}
            \includegraphics[width=0.48\textwidth]{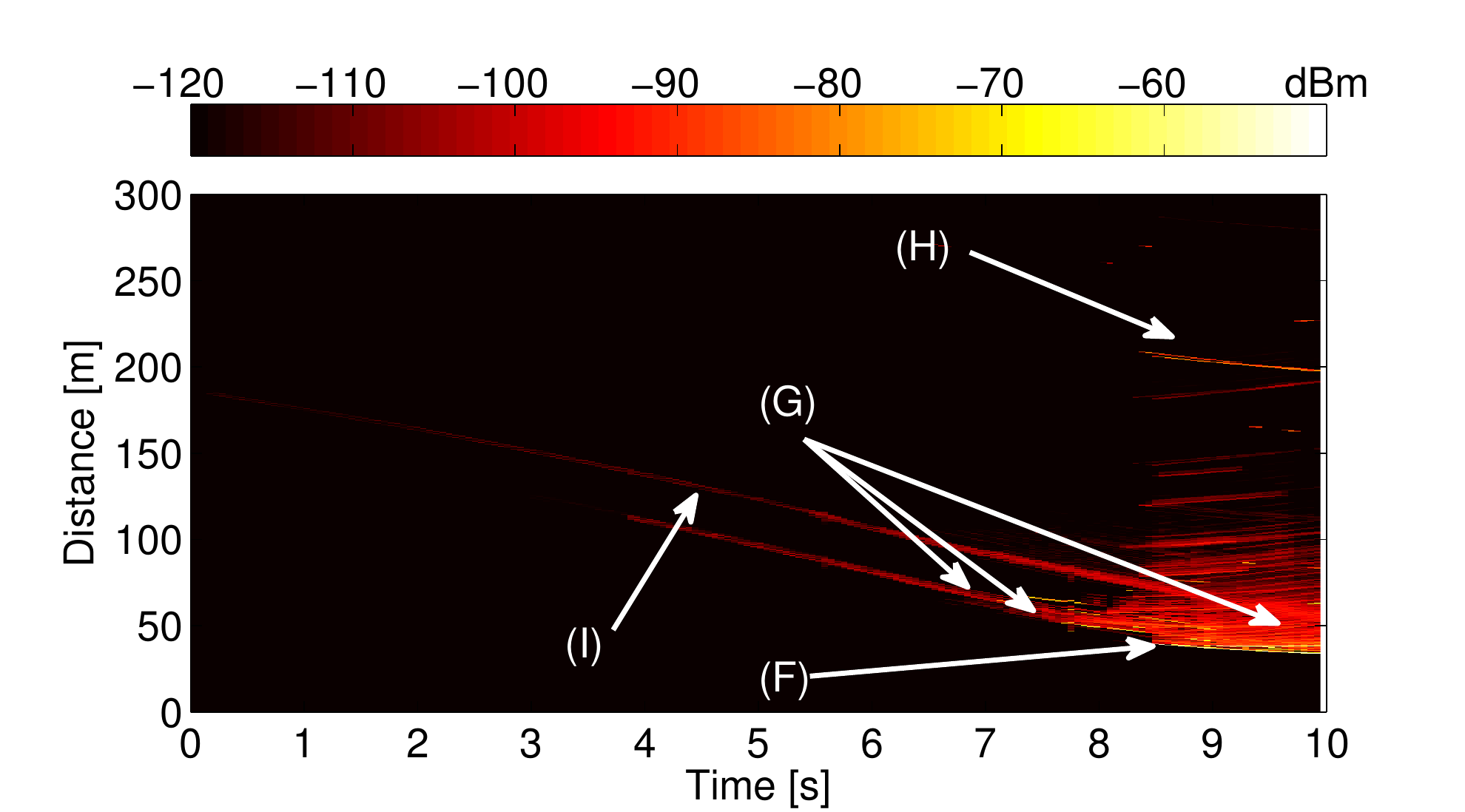}
        }%
		\\
		\subfigure[]{%
            \label{fig:Pathloss_17}
            \includegraphics[width=0.48\textwidth]{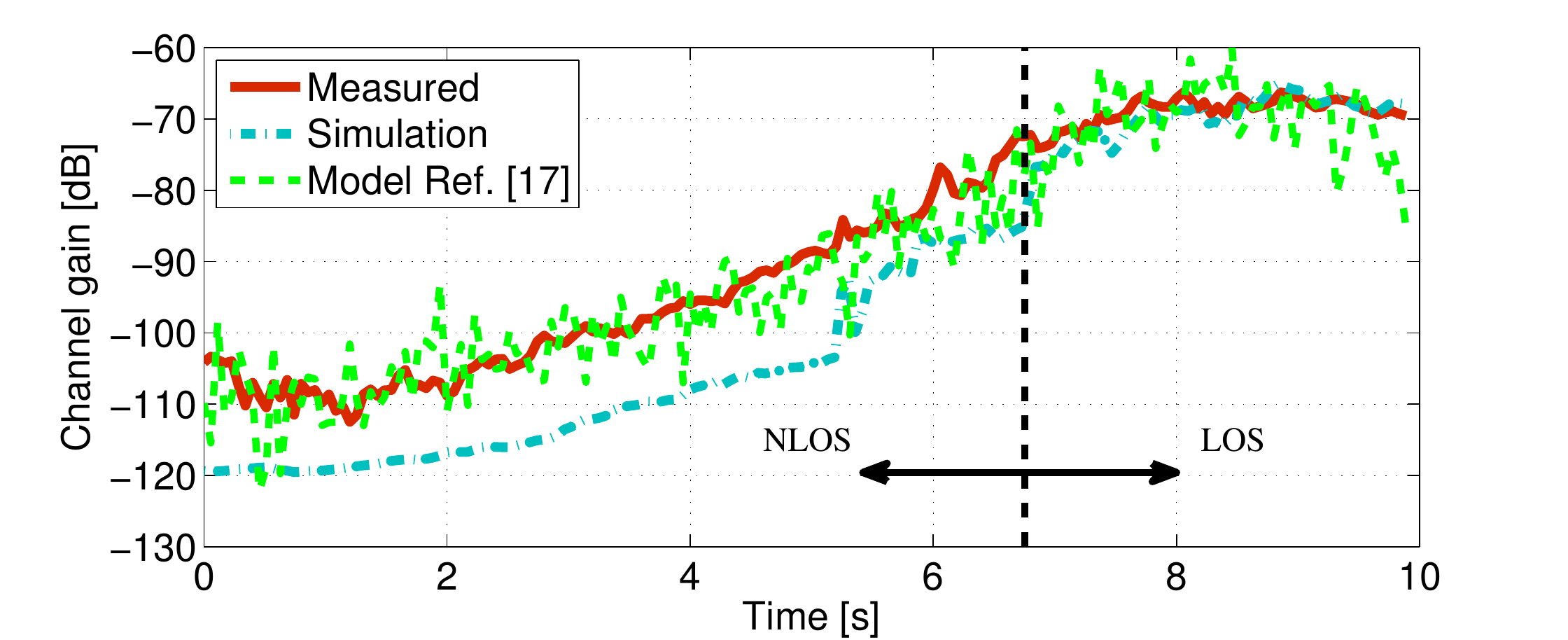}
		}%
		\subfigure[]{%
            \label{fig:Pathloss_19}
            \includegraphics[width=0.48\textwidth]{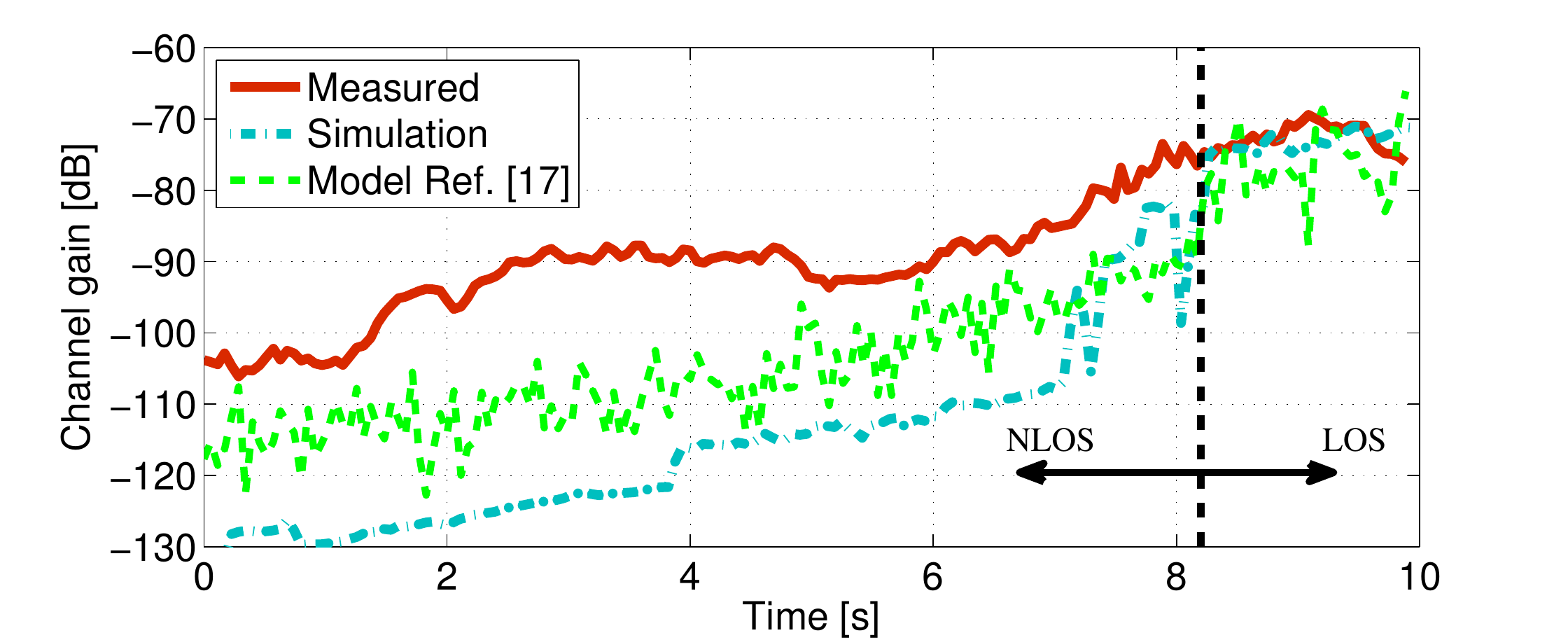}
		}%
    \end{center}
    \caption{%
         Averaged time-varying power delay profile obtained from the channel sounder data ((a), (b)), the predicted CIRs using the ray-tracing channel model ((c), (d)), and measured versus simulated channel gain compared with an empirical channel model ((e), (f)) from $M1$ and $M2$, respectively. The first and second order reflections from the static objects are considered only.
     }%
   \label{fig:PDP_17}
\end{figure*}

The resulting APDP of the measurements and the simulations are depicted in Fig.~\ref{fig:PDP_Meas_17} and~\ref{fig:PDP_RT_17} for $M1$ and in Fig.~\ref{fig:PDP_Meas_19} and~\ref{fig:PDP_RT_19} for $M2$. At first sight, we find a good agreement when comparing the simulated APDP against the measurement data. Several MPC can be identified in both figures. However, there are individual discrete scatterers as well as diffuse scatterers that can be found in the measured APDP but not in the simulated one and vice versa. In the following, the observed differences will be discussed in more detail.

In the very first few seconds the TX and RX were far from the street intersection and the LOS between them was blocked by the buildings at the corners in both scenarios, the LOS components (C) and (F) appear at approximately $6.8$\,s and $7.9$\,s in $M1$ and $M2$, respectively. The number of MPCs originating from discrete as well as diffuse scatterers in the measured APDP is much higher than in the simulated APDP because of the incomplete building data. In the simulations, the faces of each building are assumed to be plane surfaces with a single reflection coefficient, and the effects of doors, windows and balconies have not been taken in to account. Moreover, in the simulations MPCs with first and second order reflections from the static objects are considered only. In urban environments, reflections of much higher order up to $12$ as shown in \cite{Zheda2012} exist. However, the reflections with order higher than two in LOS and four in NLOS do not have significant contribution to the received power at $5.9$\,GHz as shown in \cite{Joerg2013}.  

The group of arrows (D) and (G) in Fig.~\ref{fig:PDP_17} point at several specular and non-specular MPCs that possibly originate from nearby buildings and are captured in measurements as well as in simulations. There are power contributions (E) and (H) originating from the same building which has a metallic sheet on its walls but it does not act as perfect metallic surface in reality. However, in the simulations that building is considered as a metallic object which results in a slightly higher power contribution in contrast to the measurements for that component. This example, and other similar cases points to the need for a very detailed description of the environment if one aims for an almost ideal match between measurements and simulations. But as seen, for the considered scenario, a good match can be achieved with reasonable complexity using available building database.

\subsection{Channel Gain}
Based on the APDP, we calculate the time-variant channel gain, which includes the impact of the used antennas and the system loss, as
\begin{equation}
G(t_k)=\sum_{\tau}{P_\tau(t_k,\tau)}.
\label{eq:channelgain}
\end{equation}
However, for the measured channel gain the impact of noise has been removed by setting any component below the noise threshold plus $3$\,dB to zero in the APDPs. Similar noise thresholding has been performed when calculating the delay and Doppler spreads. The noise power is estimated from the regions where no signal is present in the APDPs. Figures~\ref{fig:Pathloss_17} and~\ref{fig:Pathloss_19} show the predicted channel gains obtained from ray-tracing simulations compared with the measurement data as well as with the empirical channel model for both the scenarios. The simulated channel gains correspond to a plain scenario, where static objects like buildings, street signs and lamp posts are included, but all the parked and moving vehicles have been removed in order to reduce the simulation time. Note that we have also compared the simulation results of the full scenario including everything mentioned above with the plain scenario and have found only a marginal differences in the channel gain values.

In Fig.~\ref{fig:Pathloss_17}, a very good agreement is found between measurement, model and simulation results in the LOS region ($t\geq6.8$\,s) and in the transition region from NLOS to LOS ($6.8$\,s$\geq t \geq 5.7$\,s). Similarly, in Fig.~\ref{fig:Pathloss_19}, a very good agreement is found between measurement, model and simulation results mainly in the LOS region ($t\geq 7.9$\,s) and the region $1$\,s before the LOS. The most significant power contribution arises from the LOS component and the first order specular and non-specular reflections that are captured by the ray-tracer. There is a noticeable difference of about $8-10$\,dB during the NLOS periods ($t \leq 5.7$\,s) and $6-40$\,dB during the NLOS periods $t \leq 7.9$\,s) in Fig.~\ref{fig:Pathloss_17} and Fig.~\ref{fig:Pathloss_19}, respectively. However, the difference between the channel gain from the empirical model and simulations is relatively small due to the fact that the model is develop based on a large amount of data set and the intersections where the measurements were taken didn't have as many metallic windows or balconies as they are in considered intersections.

It is evident from the APDPs that in the NLOS situation the ray tracer underestimates the channel gain. We have found two reasonable explanations for this huge gap in NLOS situations of both scenarios: 1) The buildings in the four corners of the investigated intersections feature a lot of glass or metallic surfaces (e.g. windows and balcony elements) that yield strong power contributions at the receiver caused by specular and non-specular reflections. Especially, in scenario $M2$ we find a significant reflection (cf. arrow (I) in Fig. \ref{fig:PDP_Meas_19}) that is present even at the beginning of the scenario where both TX and RX are far away from the intersection. The same measurement was analyzed in \cite{Abbas2013ITST}, where the results were compared against a measurement based NLOS pathloss model. It was found that the NLOS model too was unable to capture such a strong reflection coming from metallic surface on the wall which is not typical for NLOS situations in urban intersections. Please note that a similar strong reflection can be identified in scenario $M1$ (cf. arrow (A) in Fig. \ref{fig:PDP_Meas_17}) which is blocked by earlier mentioned the left-turning bus for $0.5$\,s $< t \leq 3$\,s. The corresponding MPCs can also be found in the simulation-based PDPs (cf. arrows (B) and (I) in Fig. \ref{fig:PDP_RT_17} and Fig. \ref{fig:PDP_RT_19}, respectively). However, the ray tracer treats the walls of the buildings as homogeneous and concrete surfaces which explains why the power contribution in the simulations is too small. The inclusion of individual objects, like windows or balconies, with different material parameters need further implementation efforts, leading also to an increased computational complexity of the model but should of course also result in better accuracy. 

The second reasonable explanation is: 2) The ray tracer does not capture the effects of higher-order non-specular reflections at all since these phenomena are not implemented in the mathematical model. Especially, the metallic and glass element on the buildings close to the intersection provide good conditions for multi-bounce interactions of the transmitted waves on the way to the receiver. This fact leads to an additional mismatch between measurement and simulation for the investigated intersection.

The results are summarized by presenting the mean error
\begin{equation}
\mu = \text{E}\{\epsilon (t)\}
\label{eq:mean}
\end{equation} 
and the standard deviations
\begin{equation}
\sigma = \sqrt{\text{E}\left\{|\mu-\epsilon (t)|^2\right\}}
\label{eq:std}
\end{equation}
between simulated and measured channel gains where the error $\epsilon$ is given by
\begin{equation}
\epsilon (t)= G_{meas}(t)-G_{sim}(t).
\label{eq:error}
\end{equation}
The values are given in Table~\ref{tab:error} for both the measurements $M1$ and $M2$, separately for the LOS and NLOS.

\begin{figure}
     \begin{center}
        \subfigure[]{%
            \label{fig:rms_delayspread_17}
            \includegraphics[width=0.45\textwidth]{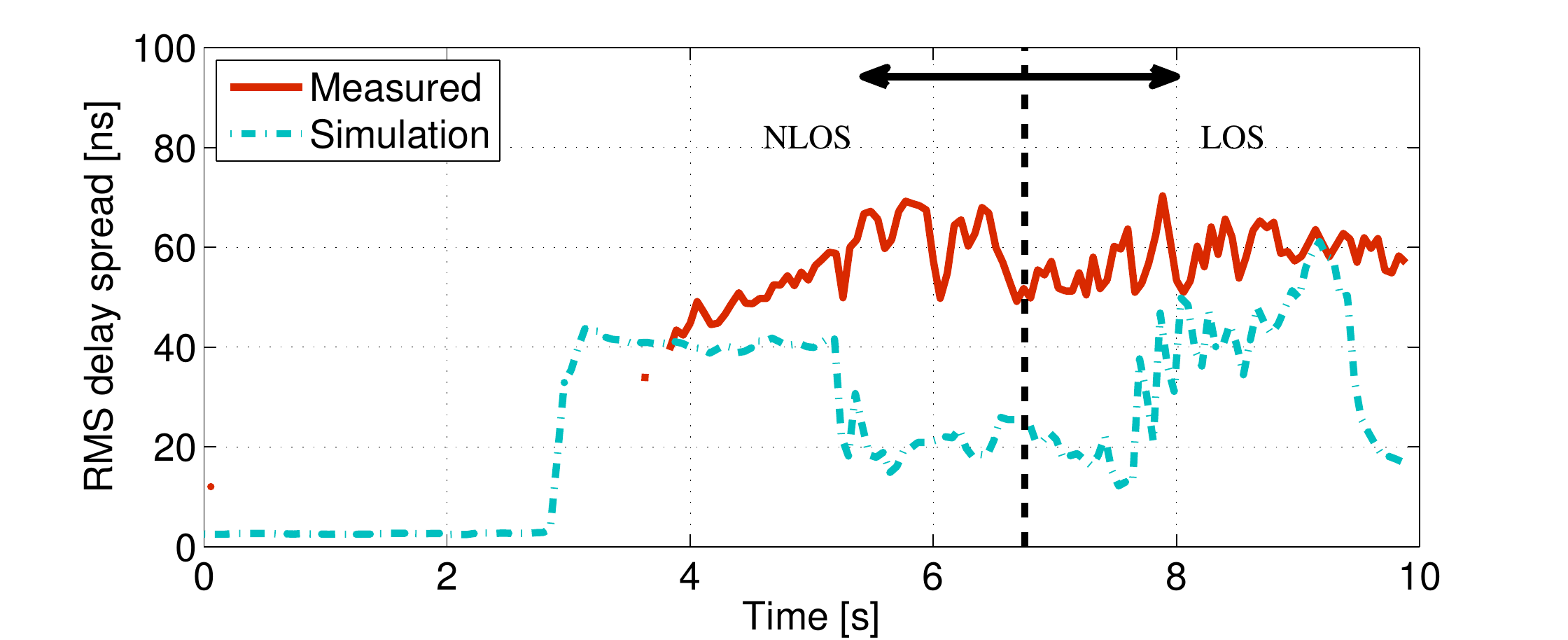}
        }%
        \\
        \subfigure[]{%
            \label{fig:rms_delayspread_19}
            \includegraphics[width=0.45\textwidth]{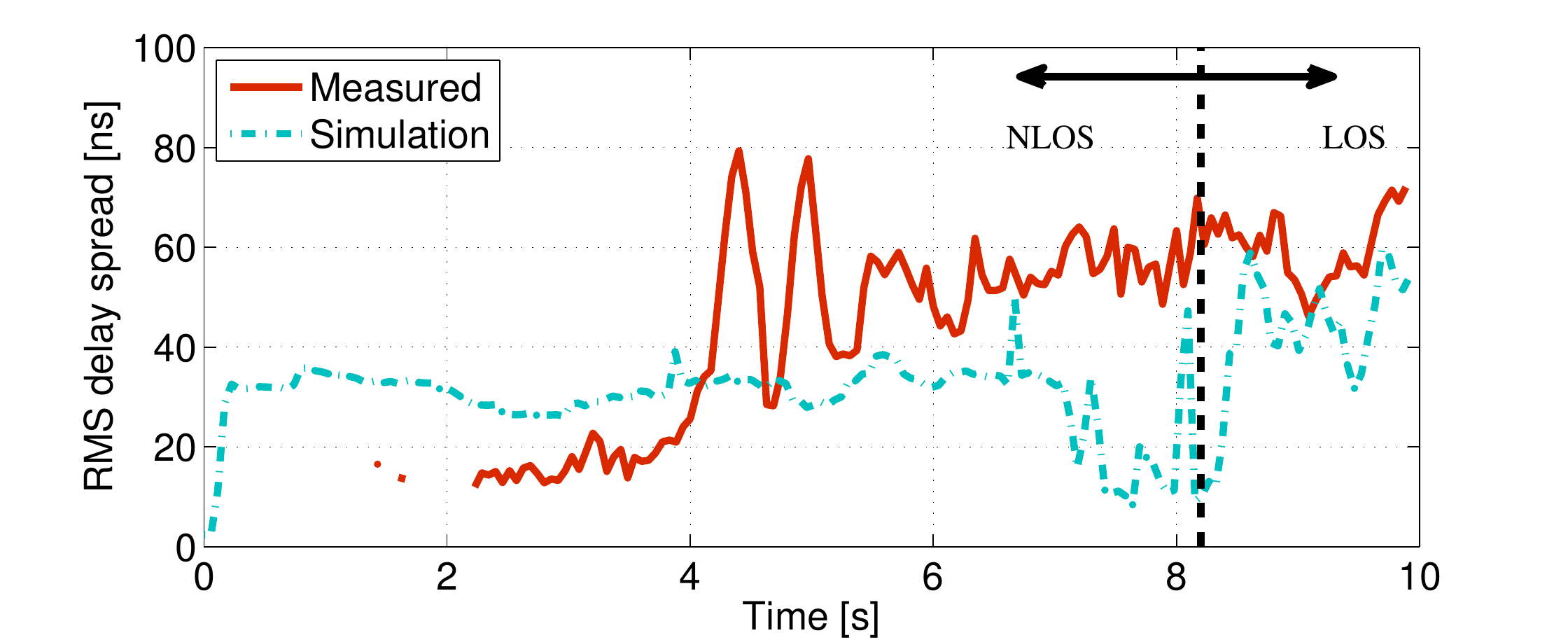}
        }%
    \end{center}
    \caption{%
        Measured versus simulated time-varying RMS delay spread of: (a) $M1$ - with percentage error of $48.2 \%$ in NLOS and $41.5 \%$ in LOS, and (b) $M2$ - with percentage error of $45.2 \%$ in NLOS and $39.5 \%$ in LOS.
     }%
   \label{fig:delayspread}
\end{figure}

\subsection{Delay and Doppler Spreads}
In a multipath propagation environment the signal spreads both in the delay and the Doppler domains. A number of delayed and scaled copies of the transmitted signal arrive at the receiver, and the effect of motion of the TX, RX or scatterers induce frequency and time selective fading that can be characterized by the root mean square (RMS) delay and Doppler spreads, respectively. These measures inversely proportional to the coherence bandwidth and coherence time of the channel, respectively \cite{Molisch1999}. The instantaneous RMS delay spread is the normalized second-order central moment of the time-variant PDP $P_\tau(t_k,\tau)$ and is defined as

\begin{equation}
S_\tau(t_k)=\sqrt{\frac{\sum_{i}P_\tau(t_k,\tau_i)\tau_i^2 }{\sum_{i}P_\tau(t_k,\tau_i)}-\left(\frac{\sum_{i}P_\tau(t_k,\tau_i)\tau_i }{\sum_{i}P_\tau(t_k,\tau_i)}\right)^2}.
\label{eq:rms_delayspread}
\end{equation}
The measured and simulated RMS delay spreads are shown in Fig.~\ref{fig:rms_delayspread_17} for $M1$ and in Fig.~\ref{fig:rms_delayspread_19} for $M2$. The delay spread of the simulated data is mostly smaller than the delay spread of measured data because the number of specular MPCs captured by the ray tracer are much smaller than of the measurements due to the limited information about the scattering environment. In addition to the aforementioned discrete MPCs that are missing, parts of the diffuse scattering that can be observed in the measured PDP with large delays do not appear in the simulations. The mean error and standard deviation are obtained by (\ref{eq:mean}) and (\ref{eq:std}), respectively, whereas $\epsilon (t)$ refers to the error between the measured and simulated RMS delay spread as a function of time. The values are summarized in Table~\ref{tab:error}. We find a reasonable agreement, with a mean error of around $24$\,ns mean error in the LOS and approximately the same in the NLOS situation for both $M1$ and $M2$. As a result, we notice the capability of the deterministic channel model to provide reliable information about the time-dispersive behavior of the urban channel, though with some underestimation

\begin{figure}
     \begin{center}
        \subfigure[]{%
            \label{fig:rms_Dopplerspread_17}
            \includegraphics[width=0.45\textwidth]{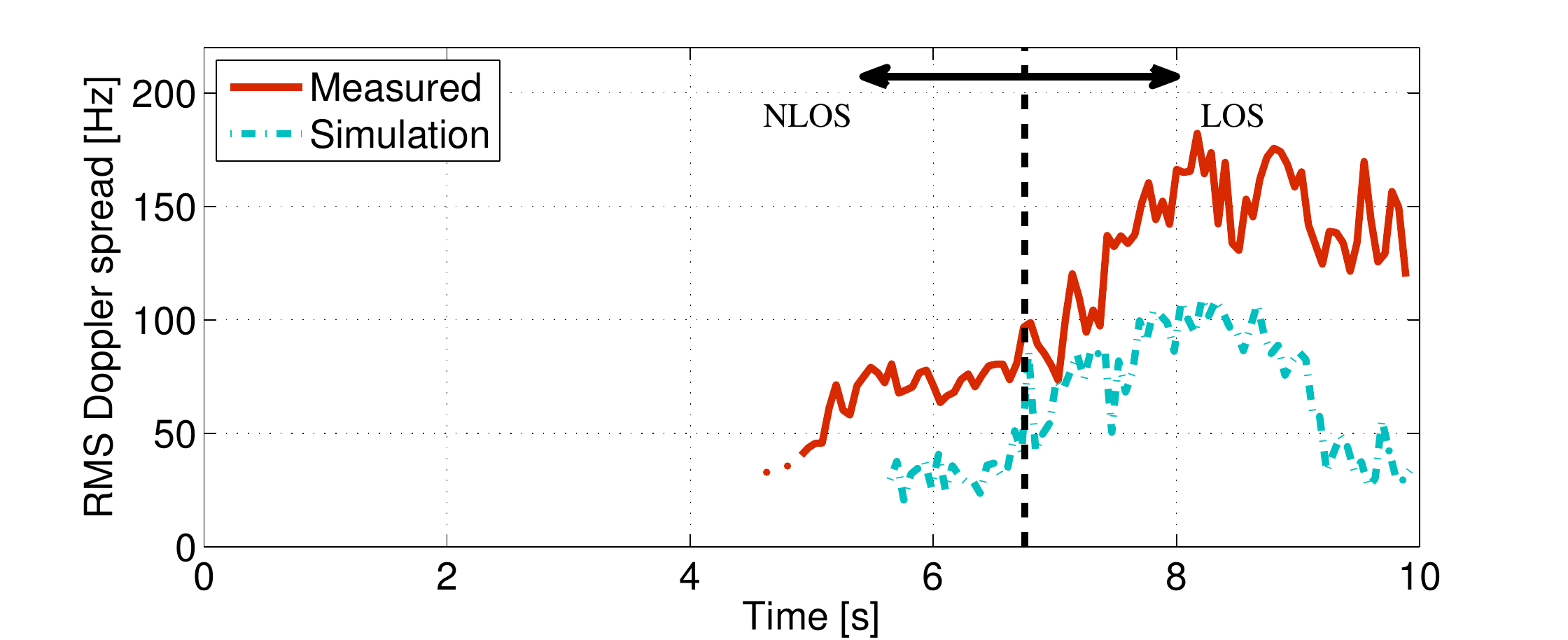}
        }%
        \\
        \subfigure[]{%
            \label{fig:rms_Dopplerspread_19}
            \includegraphics[width=0.45\textwidth]{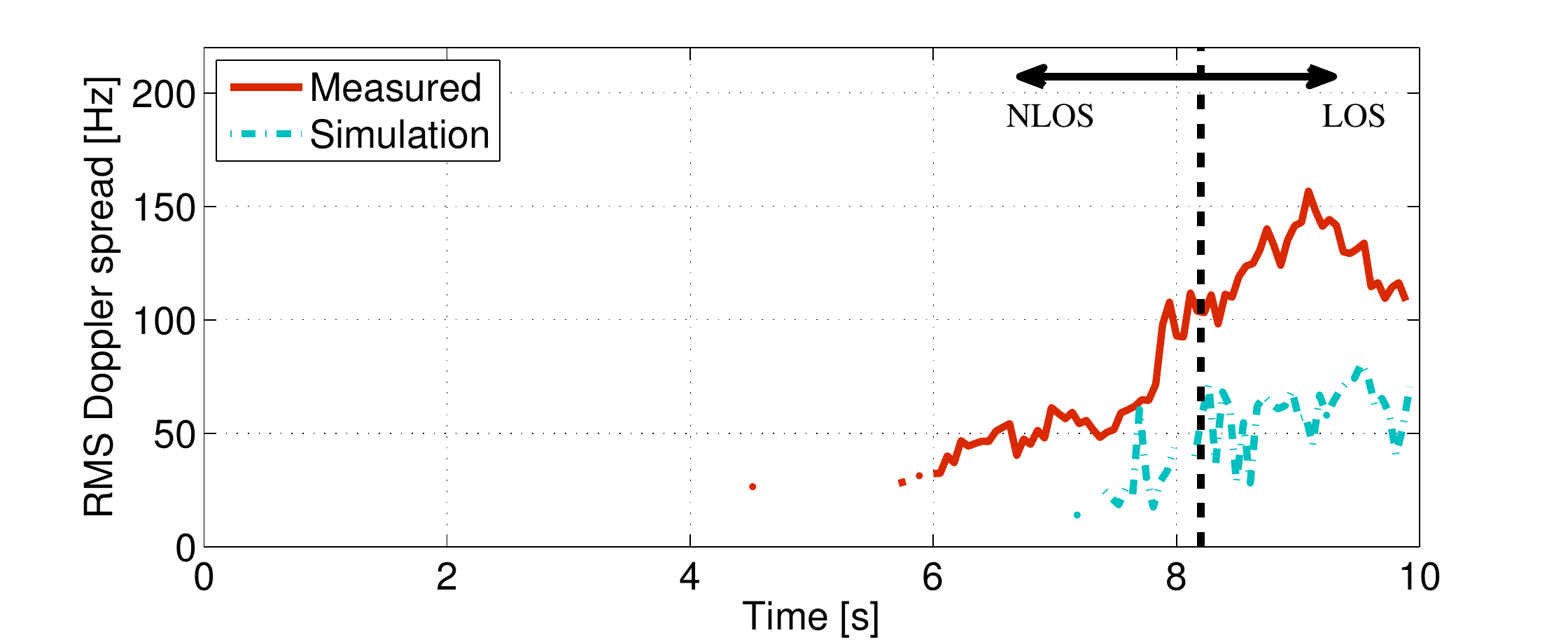}
        }%
    \end{center}
    \caption{%
        Measured versus simulated time-varying RMS Doppler spread of: (a) $M1$ - with percentage error of $52.4 \%$ in NLOS and $47.2 \%$ in LOS, and (b) $M2$ - with percentage error of $62.4 \%$ in NLOS and $53.9 \%$ in LOS..
     }%
   \label{fig:Dopplerspread}
\end{figure}

Similarly, the instantaneous RMS Doppler spread is the normalized second-order central moment of the time-variant Doppler spectral density (DSD). The DSD $P_\nu(t_k,\nu)$ is calculated analogous to PDP by taking the Fourier transform in the time domain over a sliding window of $N_{avg}$ snapshots and averaged over the delay domain. The Doppler spread can be computed as

\begin{equation}
S_\nu(t_k)=\sqrt{\frac{\sum_{i}P_\nu(t_k,\nu_i)\nu_i^2}{\sum_{i}P_\nu(t_k,\nu_i)}-\left(\frac{\sum_{i}P_\nu(t_k,\nu_i)\nu_i}{\sum_{i}P_\nu(t_k,\nu_i)}\right)^2}.
\label{eq:rms_Dopplerspread}
\end{equation}
The measured and simulated RMS Doppler spreads are shown in Fig.~\ref{fig:rms_Dopplerspread_17} for $M1$ and in Fig.~\ref{fig:rms_Dopplerspread_19} for $M2$. The mean error and standard deviation are obtained by (\ref{eq:mean}) and (\ref{eq:std}), respectively, whereas $\epsilon (t)$ here refers to the error between the measured and simulated RMS Doppler spread and are summarized in Table~\ref{tab:error}. As mentioned above, only static objects are considered in the simulations and the number of MPCs captured by the ray tracer is smaller than what can be identified from the measurements. Although the Doppler spread of the simulated data is smaller than that of the measured data, the ray tracing model is able to capture most of the significant MPCs. The higher Doppler spread in the measured data is due to many incoming waves from all directions (angles) at the receiver, originating from scatterers such as moving and parked vehicles, details at the building walls, etc. The ray tracing model does not capture all of these details, in turn results in a smaller spread in the Doppler domain.


\subsection{Antenna Diversity}

Random fluctuations in the signal power due to multipath propagation caused by scattering impair the wireless channel across space, time or frequency. This is commonly known as channel fading. Diversity techniques are developed to combat fading by combining several independently faded version of the same transmitted signal at the receiver to improve link reliability by improving the signal-to-noise ratio (SNR). Thus, diversity is an important metric to be analyzed in order to validate the performance of ray-tracing channel simulations for MIMO V2V systems.

Among the diversity techniques, here spatial, or antenna, diversity is of particular interest in which multiple antennas at the TX and/or RX are used to exploit the diversity gain. In order to evaluate antenna diversity captured in the ray-tracing simulations we compare two metrics, the eigenvalue distribution and antenna correlation in the following.
\begin{figure*}
     \begin{center}
        \subfigure[]{%
            \label{fig:eigenvalue_17}
            \includegraphics[width=0.40\textwidth]{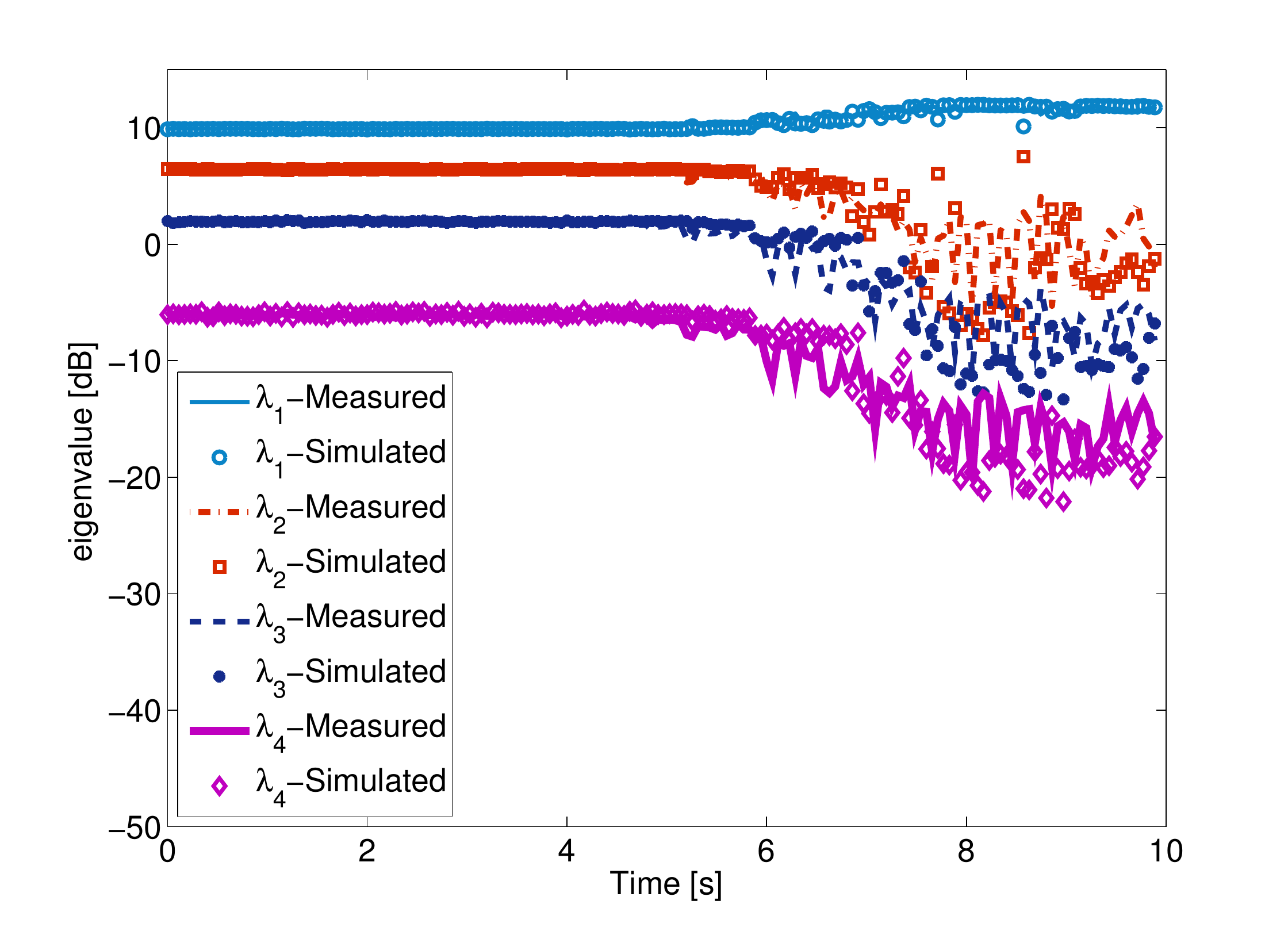}
        }%
        \subfigure[]{%
            \label{fig:eigenvalue_19}
            \includegraphics[width=0.40\textwidth]{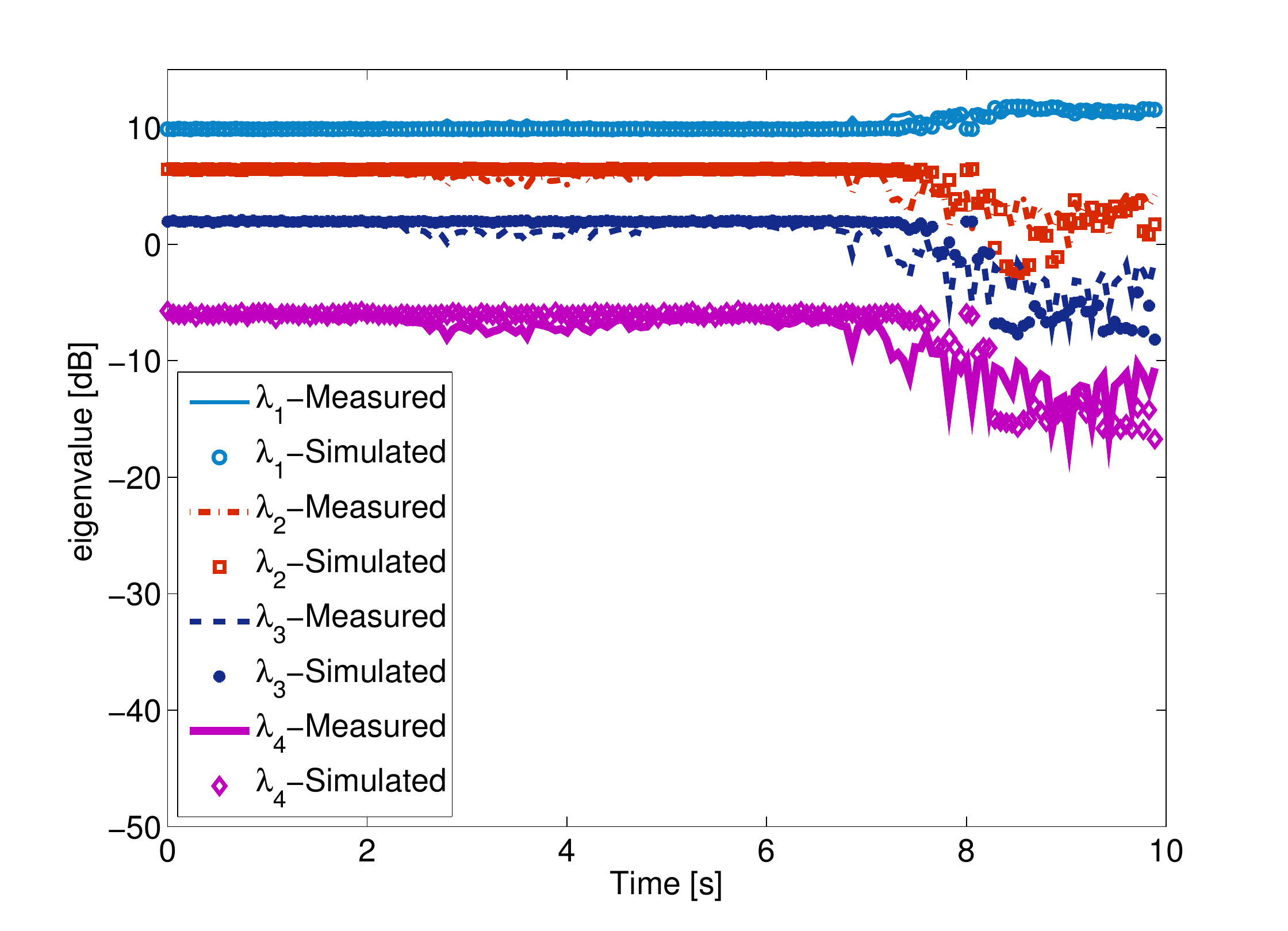}
        }%
				\\
				\subfigure[]{%
            \label{fig:eigenvalue_meas_17}
            \includegraphics[width=0.40\textwidth]{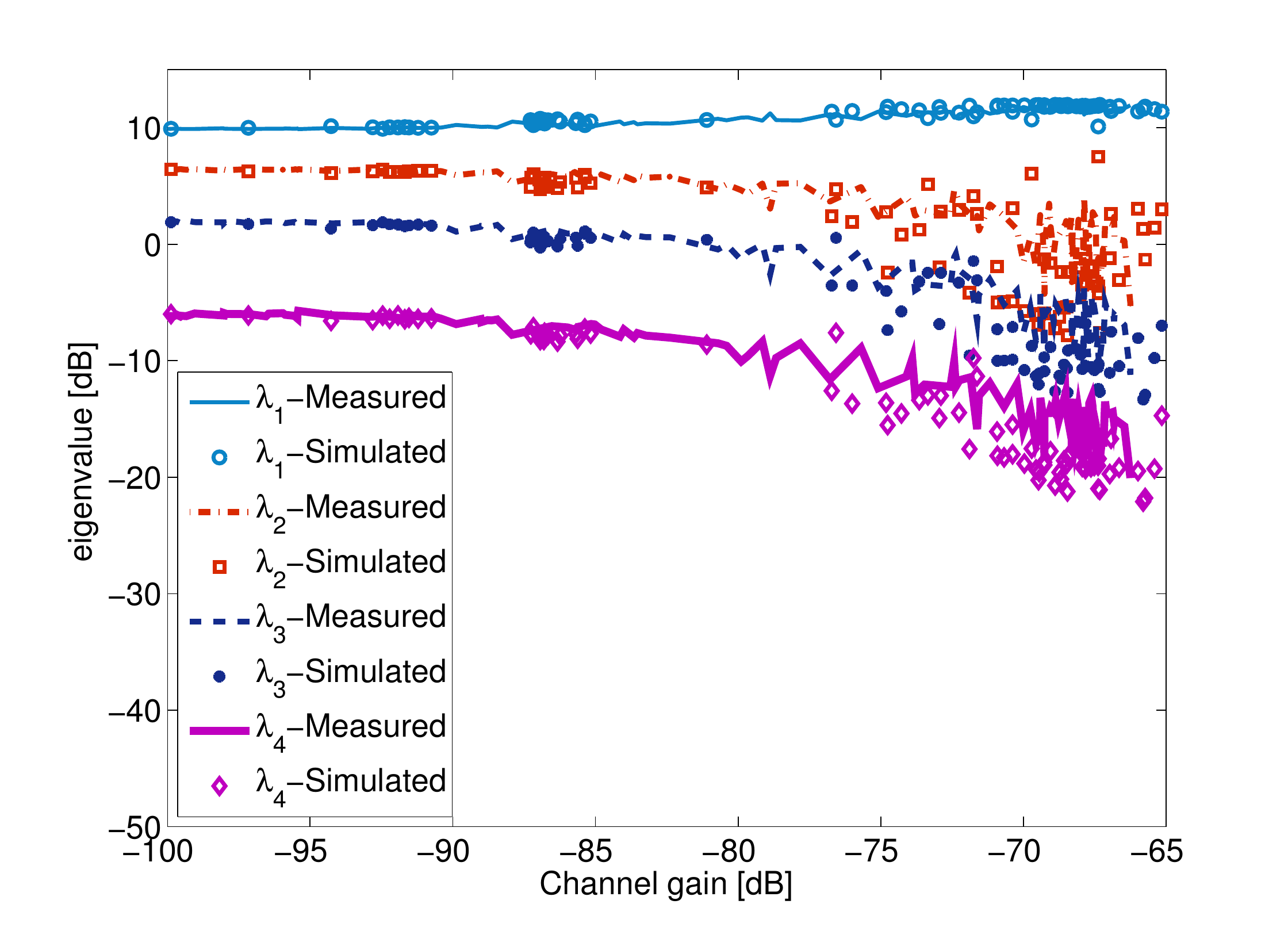}
        }%
        \subfigure[]{%
            \label{fig:eigenvalue_meas_19}
            \includegraphics[width=0.40\textwidth]{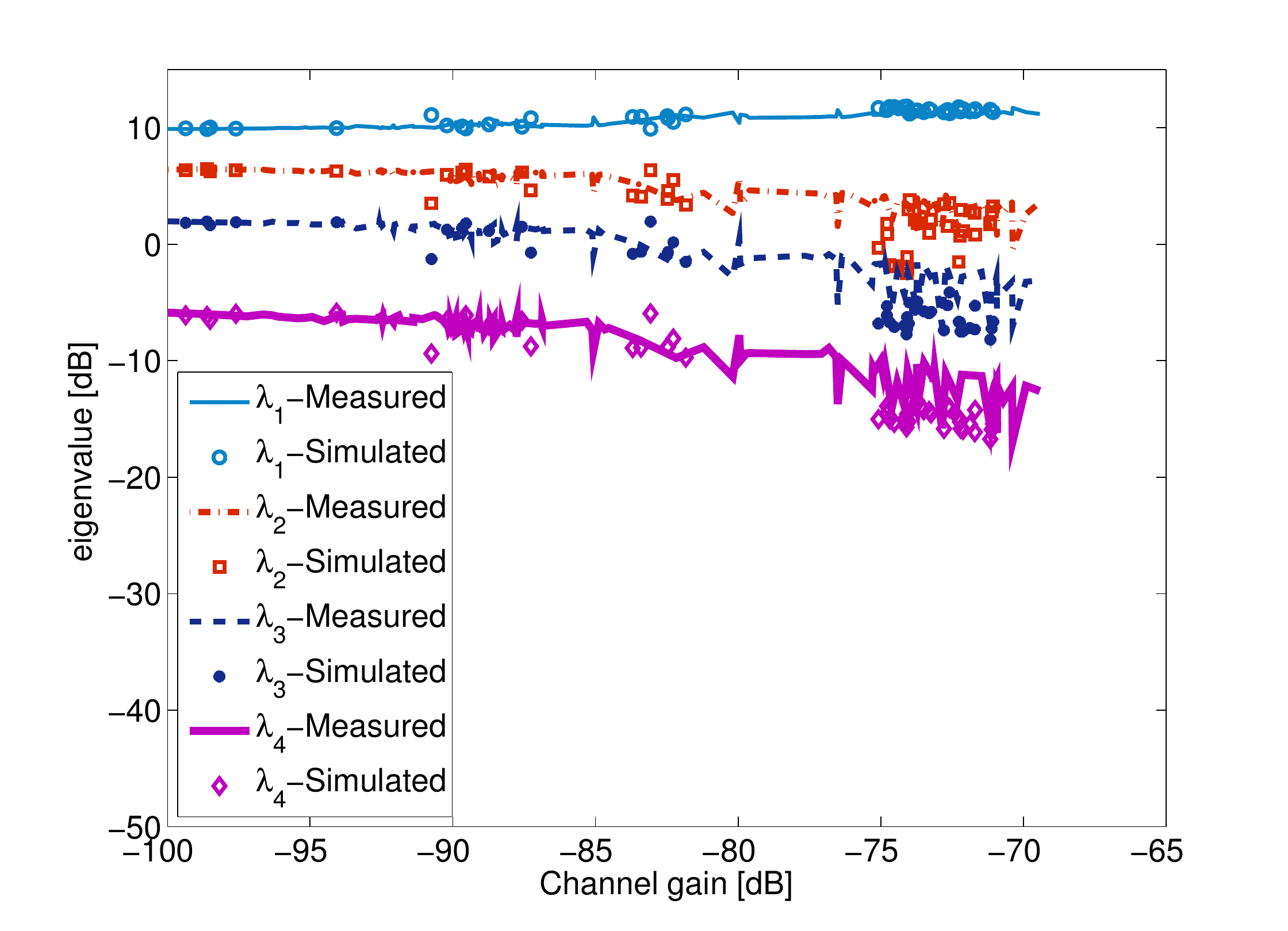}
        }%
    \end{center}
    \caption{%
		Measured versus simulated eigenvalues: Figs.~\ref{fig:eigenvalue_17}, and \ref{fig:eigenvalue_meas_17} represent $M1$ and Figs.~\ref{fig:eigenvalue_19}, and \ref{fig:eigenvalue_meas_19} represent $M2$.
     }%
   \label{fig:eigenvalues}
\end{figure*}

\begin{figure*}
     \begin{center}
        \subfigure[]{%
            \label{fig:Ant_corr_12_17}
            \includegraphics[width=.45\textwidth]{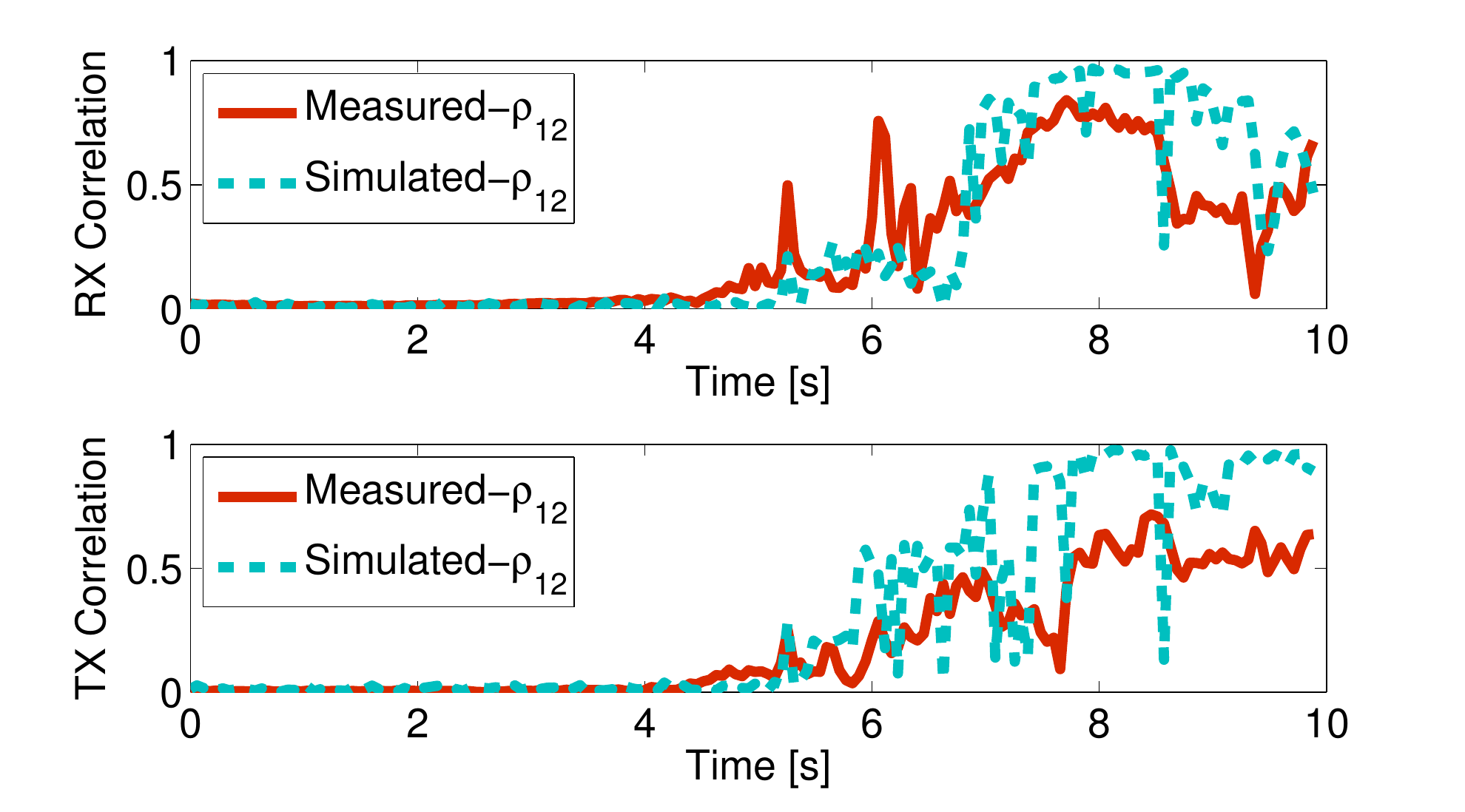}
        }%
				 \subfigure[]{%
            \label{fig:Ant_corr_12_19}
            \includegraphics[width=.45\textwidth]{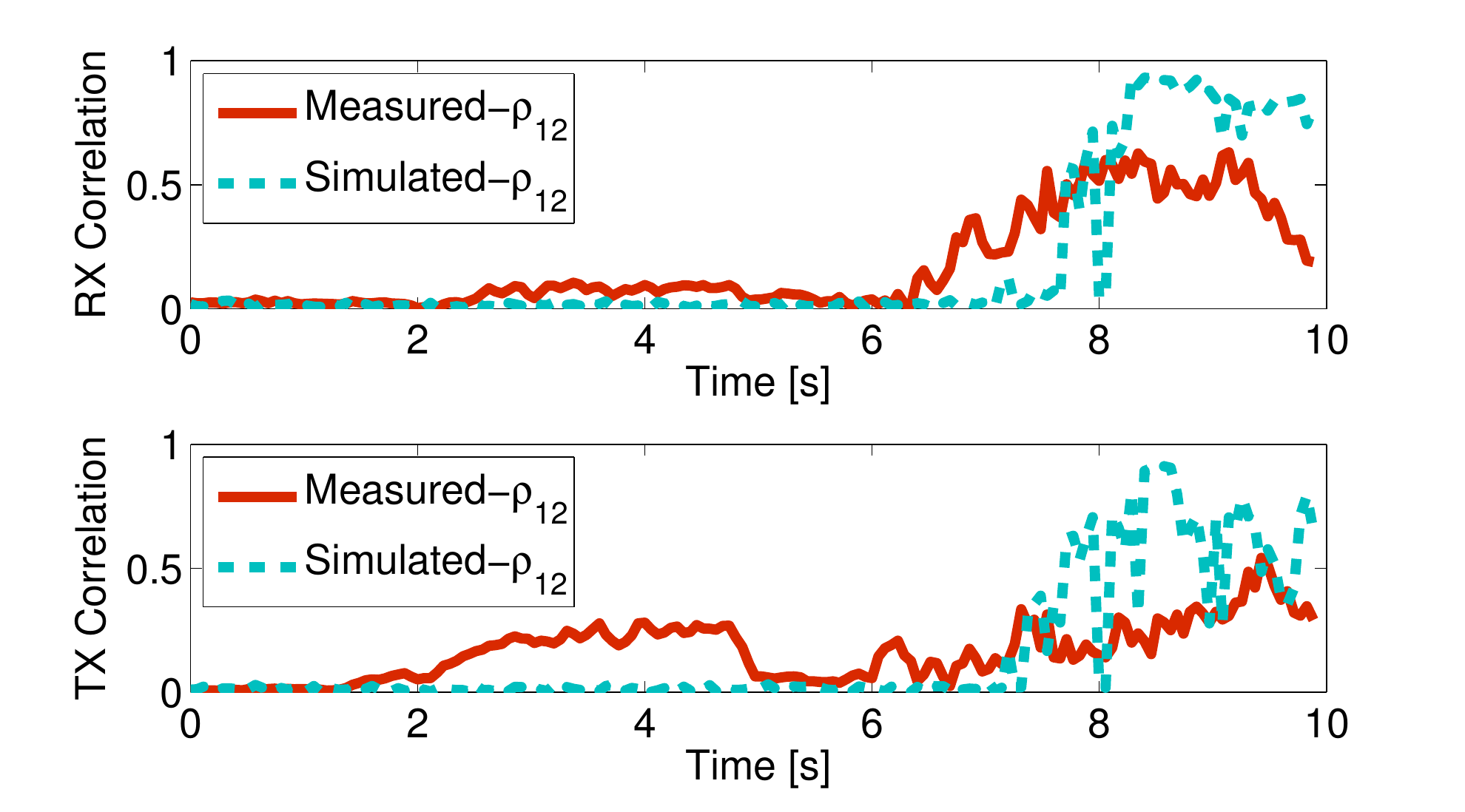}
        }%
  			\\ 
				\subfigure[]{%
            \label{fig:Ant_corr_23_17}
        		\includegraphics[width=.45\textwidth]{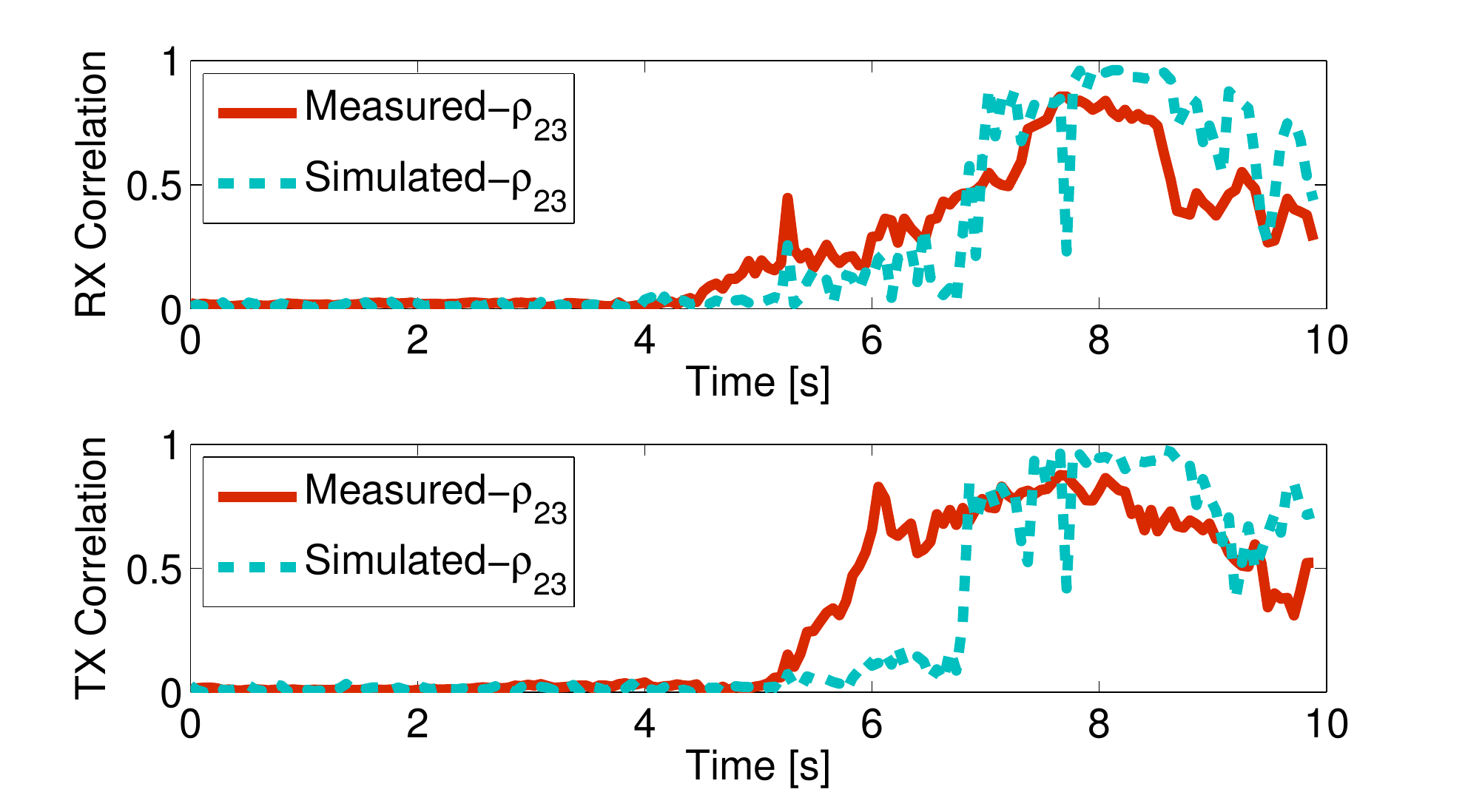}  			
  			}
  			 \subfigure[]{%
            \label{fig:Ant_corr_23_19}
        		\includegraphics[width=.45\textwidth]{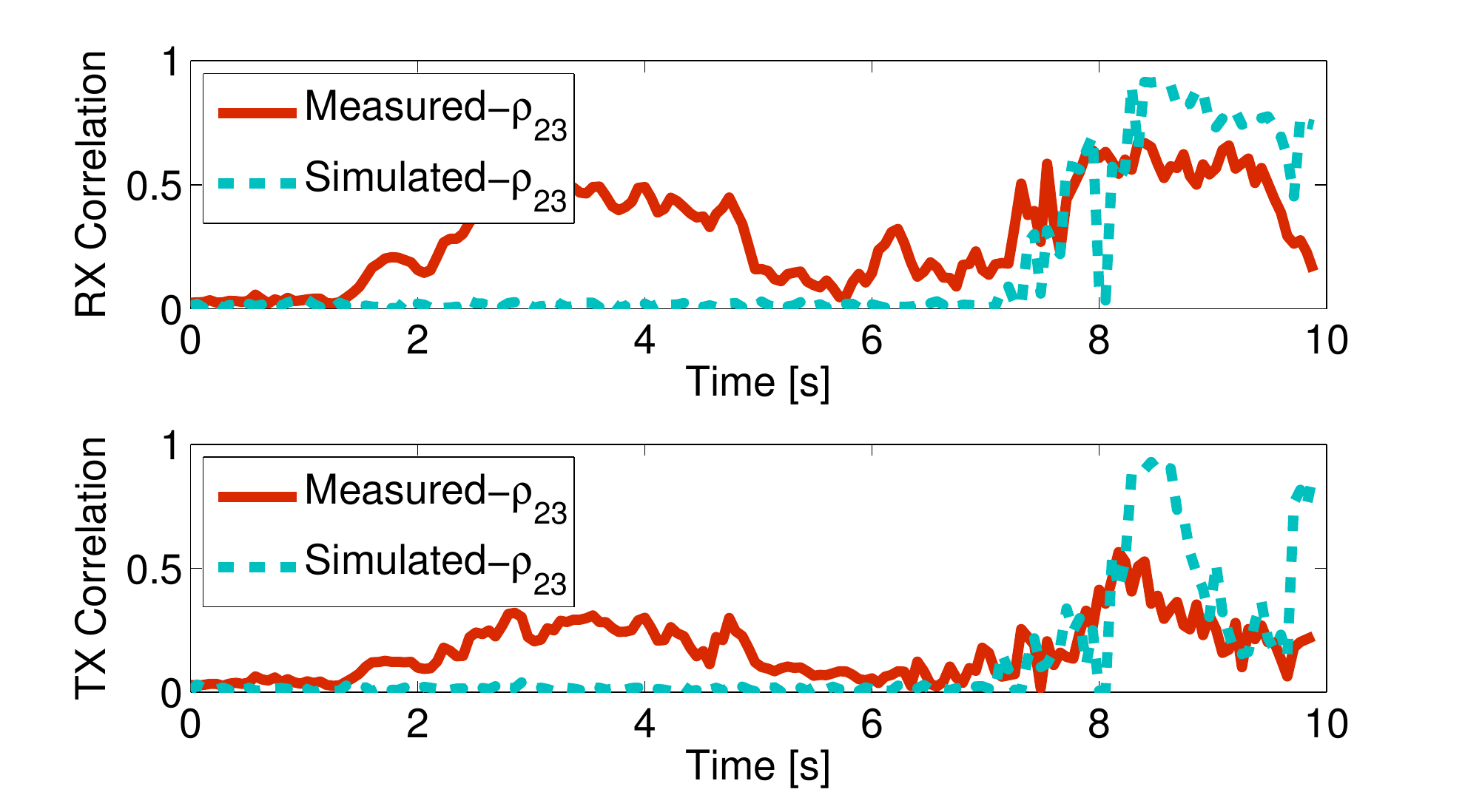}
  			} 
        \\ 
				\subfigure[]{%
  					\label{fig:Ant_corr_34_17} 
    				\includegraphics[width=.45\textwidth]{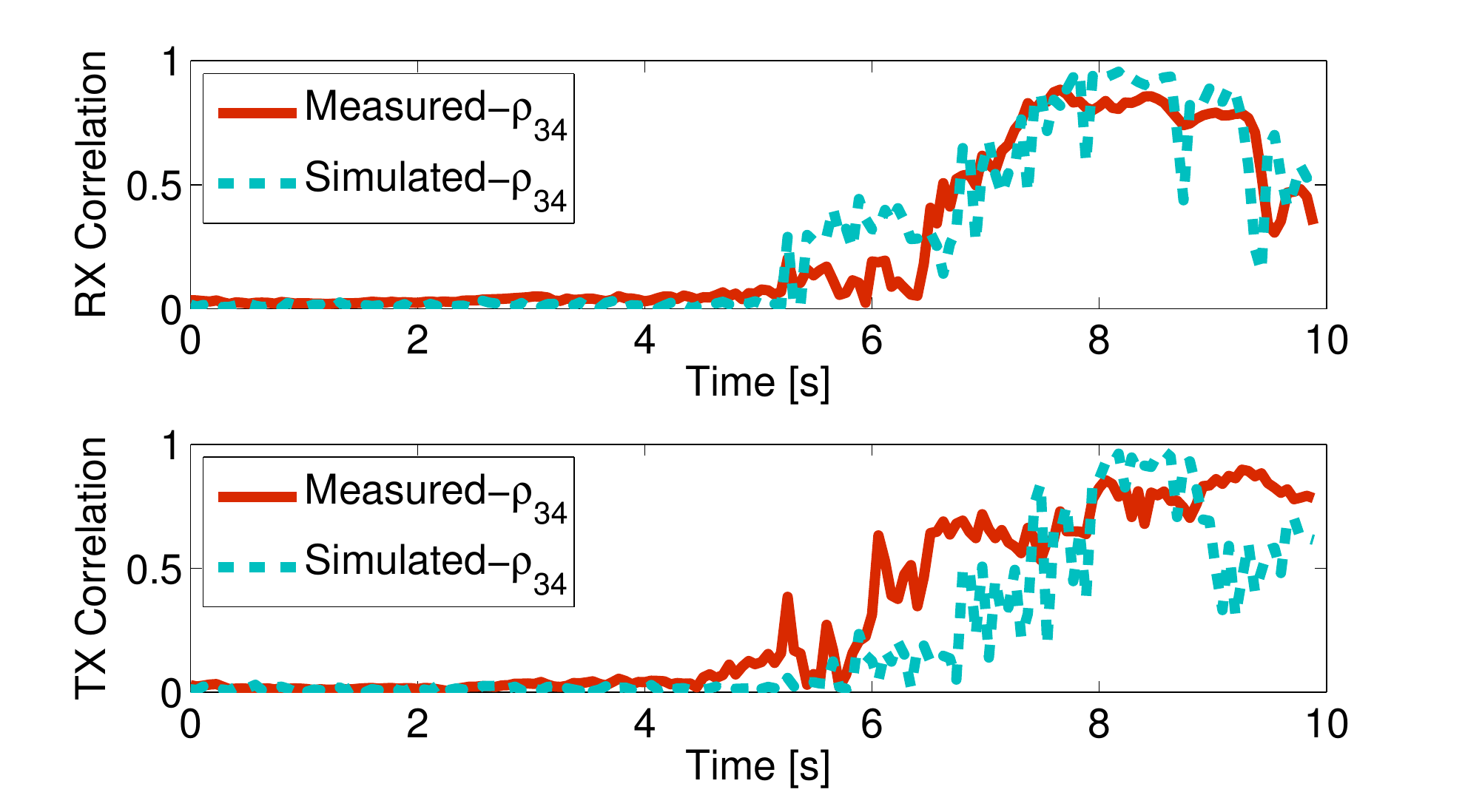}
  			}  
  			\subfigure[]{%
  					\label{fig:Ant_corr_34_19} 
    				\includegraphics[width=.45\textwidth]{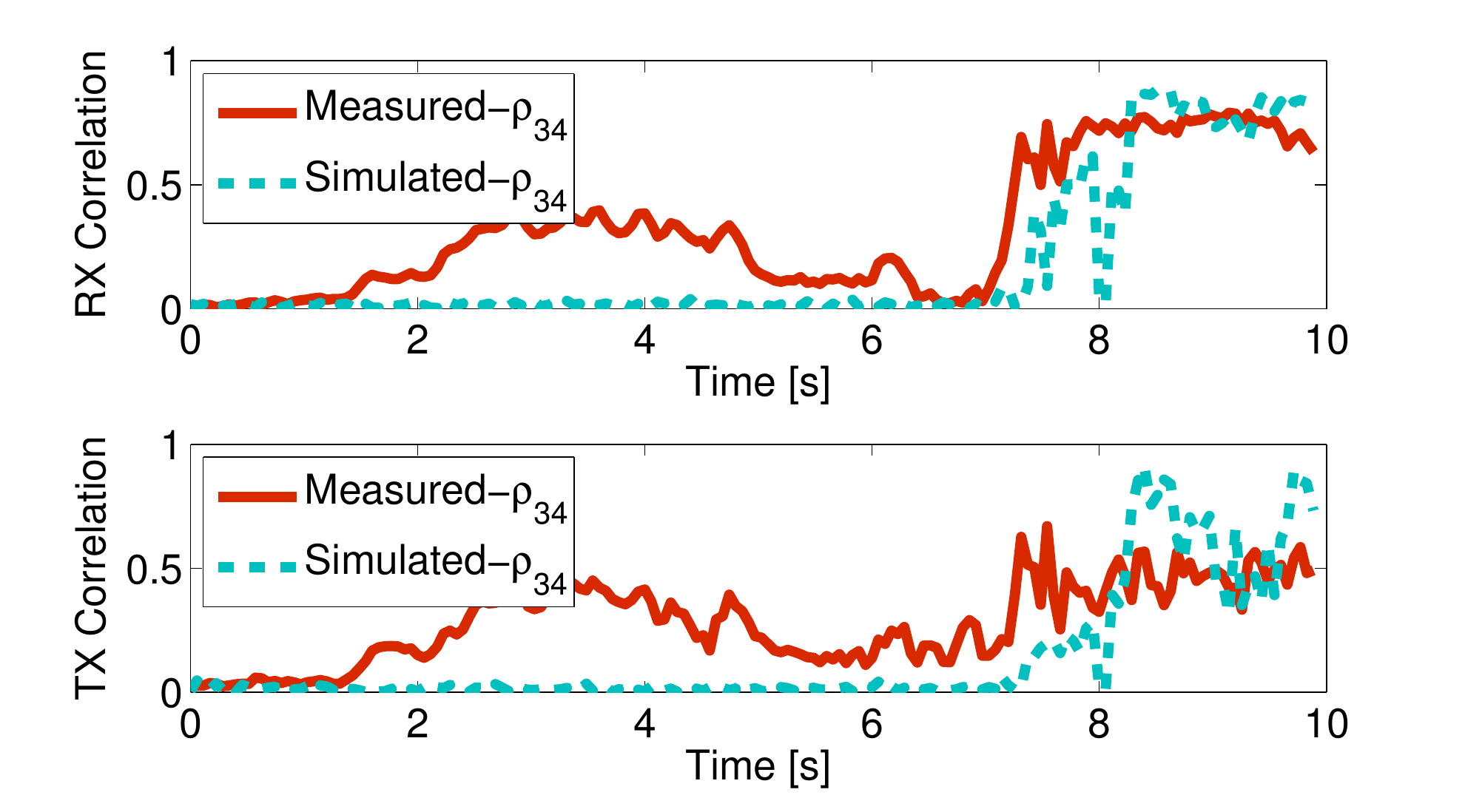}
  			}
    \end{center}
    \caption{The correlation between the antenna elements $1-2$, $1-3$, and $1-4$ is shown as a function of time. Figs.~\ref{fig:Ant_corr_12_17},~\ref{fig:Ant_corr_23_17}, and~\ref{fig:Ant_corr_34_17} represent $M1$ and Figs.~\ref{fig:Ant_corr_12_19},~\ref{fig:Ant_corr_23_19}, and~\ref{fig:Ant_corr_34_19} represent $M2$.}
    \label{fig:Ant_corr}
\end{figure*}


\subsubsection{Eigenvalue distribution and array gain}

The eigenvalues (EVs) and their distributions obtained from ray-tracing simulations are compared against the ones obtained from the measurement data as they capture important properties of the array and the medium \cite{Vaughan2003}. An SVD expansion of the normalized channel matrix $H\in\mathbb{C}^{M_R \times M_T}$ can be written as,

\begin{equation}
H = U\cdot S\cdot V^*,
\label{SVD}
\end{equation}
where $U$ is an $M_R\times M_R$ unitary matrix, $S$ is a diagonal matrix of real non-negative singular values $\sigma_m$ where $m=1,2,...,\min\{M_R,M_T\}$, and $V^*$ (the conjugate transpose of $V$) is an $M_T\times M_T$ unitary matrix. Singular values of $H$ are the square roots of the eigenvalues of $H \cdot H^*$.

In Fig.~\ref{fig:eigenvalue_17} and~\ref{fig:eigenvalue_19}, the time evolution of the eigenvalues, $\lambda_m$ where $m=1,2,...,\min\{M_R,M_T\}$, of the measured and simulated data are compared for both $M1$ and $M2$. The markers represent measured eigenvalues whereas the lines represent simulated eigenvalues.  It is interesting to notice that the eigenvalues of the measured and simulated data are very similar. For the first $5$\,s in both $M1$ and $M2$ the eigenvalues look the same, but the scenario is NLOS and the available signal power is very weak. In such a situation the noise, which is approximately i.i.d. Gaussian in both the measurements as well as in the simulations, is dominating. This in turn gives the same eigenvalues. However, in Fig.~\ref{fig:eigenvalue_19} the are some differences in the eigenvalues between $2-5$\,s because of the presence of the dominating MPC in the NLOS in $M2$ as discussed before.

In Fig.~\ref{fig:eigenvalue_meas_17}-\ref{fig:eigenvalue_meas_19} the eigenvalues are plotted as a function of channel gain. The eigenvalues are shown only for the samples where the channel gain is higher than $-100$\,dB, for the region in time where signal is dominating the noise. As the ray tracing simulations do not extract all MPCs and thus provides lower channel gain, we see some gaps in the plots where the eigenvalues correspond to lower channel gain than in reality. The mean and standard deviation of the error in the eigenvalues are listed in Table~\ref{tab:error}.
\begin{table*}
  \centering
  \caption{Mean $\mu$ and standard deviation $\sigma$ of the error in the estimated parameters obtained from the simulated data with respect to measurement data for both $M1$ and $M2$.}
    \begin{tabular}{ll|c|c|c|c}
    \toprule
    \bf{Parameters}    & & \multicolumn{2}{ c| }{\bf{LOS}} & \multicolumn{2}{ c }{\bf{NLOS}} \\
		\hline
																			&											& \bf{M1} 							& \bf{M2} 							& \bf{M1} 								& \bf{M2} \\
		\hline
																			&   									& ($\mu_1$, $\sigma_1$) & ($\mu_2$, $\sigma_1)$ & ($\mu'_1$, $\sigma'_1$) & ($\mu'_2$, $\sigma'_1$) \\
		\hline
		Channel gain 		 									& $G_h$ [dB]         	& (0.87, 1.66) & (3.12, 5.43) & (11.2, 2.75) & (26.2, 6)    \\
		\hline
		Delay spread  										& $\tau$  [ns]    	    & (24.5, 13.5) & (23.2, 17.2) & (24, 17.5) & (9.3, 20)    \\
		\hline
	  Doppler spread										& $\mu$  [Hz]        	& (64.8, 28.9) & (62.7,20.1) & (40.4, 10.2) & (34.4,6.34)    \\
		\hline
																			&											&								&							&							&							\\
																			& $\lambda_1$ [dB] 		& (-0.07, 0.38) & (-0.07, 0.49) & (0.104, 0.19) & (0.28, 0.26)  \\
		Eigenvalues												& $\lambda_2$ [dB]		& (1.88, 4.26) & (0.92, 2.5) & (-0.21, 0.51) & (-0.49, 0.59)  \\
																			& $\lambda_3$ [dB] 		& (2.41, 3.36) & (1.55, 3.18) & (-0.37, 0.76) & (-0.69, 0.75)  \\
																			& $\lambda_4$ [dB] 		& (2.41, 3.1) & (1.08, 3.23) & (-0.52, 0.98) & (-0.77, 0.82)  \\
																			&											&								&							&							&							\\
	  \hline
																			&											&								&							&							&							\\
																			& $\rho_{12}$ 				& (-0.31, 0.23) & (-0.31, 0.24) & (-0.032, 0.11) & (0.1, 0.1)  \\
		TX																& $\rho_{13}$ 				& (-0.12, 0.3) & (-0.11, 0.33) & (0.07, 0.108) & (0.117, 0.12)  \\
		antenna														& $\rho_{14}$ 				& (-0.12, 0.23) & (-0.14, 0.26) & (0.02, 0.106) & (0.17, 0.13)  \\
		correlations											& $\rho_{23}$        	& (-0.12, 0.17) & (-0.18, 0.27) & (0.1, 0.19) & (0.12, 0.1)  \\
		coefficients											& $\rho_{24}$ 				& (-0.13, 0.27) & (-0.14, 0.25) & (0.03, 0.12) & (0.17, 0.14)  \\
																			& $\rho_{34}$ 				& (0.11, 0.23)  & (-0.07, 0.23) & (0.07, 0.14) & (0.21, 0.13)  \\
																			&											&								&							&							&							\\
		\hline
																			&											&								&							&							&							\\
																			& $\rho_{12}$ 				& (-0.22, 0.21) & (-0.27, 0.27) & (0.043, 0.11) & (0.07, 0.09)  \\
		RX																& $\rho_{13}$ 				& (-0.20, 0.33) & (0.02, 0.2) & (-0.002, 0.104) & (0.04, 0.07)  \\
		antenna														& $\rho_{14}$ 				& (-0.03, 0.19) & (0.026, 0.15) & (-0.007, 0.08) & (0.06, 0.09)  \\
		correlations											& $\rho_{23}$ 				& (-0.18, 0.18) & (-0.17, 0.23) & (0.05, 0.09) & (0.21, 0.16)  \\
		coefficients											& $\rho_{24}$ 				& (0.008, 0.33) & (-0.21, 0.33) & (0.02, 0.09) & (0.16, 0.15)  \\
																			& $\rho_{34}$ 				& (-0.01, 0.16) & (0.03, 0.21) & (-0.01, 0.103) & (0.17, 0.14)  \\
																			&											&								&							&							&							\\
    \bottomrule
    \end{tabular}%
  \label{tab:error}%
\end{table*}%


\subsubsection{Antenna correlations}

Multiple antennas at the TX or at the RX can improve the system performance through diversity arrangements, but their benefits can only be fully utilized if the correlation between signals at different antenna elements is low \cite{molisch05}. Thus, antenna correlation for both the TX and the RX array is an important parameter to study. The time-variant antenna correlation $\rho_{ij}^{RX}(t_k)$ between the RX elements $i$ and $j$ is calculated as

\begin{equation}
\rho_{ij}^{RX}(t_k)=\sum_{n_t=1}^{N_{avg}}\sum_{n_f=1}^{N_f}\frac{\sum_{m=1}^{M_T} H_{i,m}H_{j,m}^*}{\sqrt{\sum_{m=1}^{M_T} |H_{i,m}|^2\sum_{m=1}^{M_T} |H_{j,m}|^2}}.
\label{eq:rho_RX}
\end{equation}
Similarly, the correlation $\rho_{ij}^{TX}(t_k)$ between the TX elements $i$ and $j$ is calculated as,

\begin{equation}
\rho_{ij}^{TX}(t_k)=\sum_{n_t=1}^{N_{avg}}\sum_{n_f=1}^{N_f}\frac{\sum_{n=1}^{M_R} H_{n,i}^*H_{n,j}}{\sqrt{\sum_{n=1}^{M_R} |H_{n,i}|^2\sum_{n=1}^{M_R} |H_{n,j}|^2}},
\label{eq:rho_TX}
\end{equation}
where $H_{n,m}$ is a block matrix for each time instant $t_k$, $n^{th}$ RX and $m^{th}$ TX elements, respectively, such that $H\in\mathbb{C}^{N_{avg}\times N_f}$ and $H^*$ is the conjugate transpose of $H$. $N_f$ is the number of frequency bins within measurement bandwidth.

In Fig.~\ref{fig:Ant_corr}, the correlation between the antenna elements is shown as a function of time. The correlation between the elements is almost zero from $t=0\,s$ until $t$ reaches approximately 6\,s. In this time interval, i.i.d. Gaussian noise is the dominating contribution, which result in mostly uncorrelated subchannels. In the absence of LOS, and it increases as the vehicle gets close to the intersection. The correlation between all elements is higher when there is LOS between the TX and RX in both $M1$ and $M2$. The antenna correlation coefficients for consecutive antenna elements $1-2$, $2-3$, and $3-4$ of both the RX and TX arrays for $M1$ are shown in Figs.~\ref{fig:Ant_corr_12_17},~\ref{fig:Ant_corr_23_17}, and~\ref{fig:Ant_corr_34_17}. The measured and simulated correlation coefficients are not exactly the same but show a similar trend over the time. Similarly, the antenna correlation coefficients for antenna elements $1-2$, $2-3$, and $3-4$ of both the RX and TX arrays for $M2$ are shown in Figs.~\ref{fig:Ant_corr_12_19},~\ref{fig:Ant_corr_23_19}, and~\ref{fig:Ant_corr_34_19}. For $M2$, again the correlation between $2-5$\,s is higher for measurement than that in simulations. However, in the presence of LOS the correlation of the TX and RX from the simulation and measurement show similar trend. The mean and standard deviation of the error in correlations is listed in Table~\ref{tab:error}.

\section{Conclusion}
\label{sec:conclusion}
The characterization of an urban intersection for V2V channel is non-trivial as many details of the environment may have a huge impact on the experienced radio channel. This includes material, position and alignment of buildings, street width, location and density of road side objects etc. An accurate modeling is required for safety-critical system design, which in turn requires a full understanding of these physical effects. Deterministic models such as those achieved with a ray tracer are anticipated to be good candidates to model such complex scenarios.

In this paper, we have presented an accuracy analysis of ray tracing based simulation of V2V channels in an urban intersection scenario with the help of real channel measurements. We have compared important channel metrics, namely power delay profile, channel gain, delay and Doppler spreads, eigenvalues and antenna correlations that are obtained from both channel sounder measurements and simulation results using a ray-based channel model. We have found a good accuracy of the simulation results in the sense that the present physical phenomena of wave propagation are captured by the ray tracer. Variations between measurements and simulations are consistent and can be explained with the microscopic features of the investigated scenario. The current release of the ray-tracing routines constitue a good basis to characterize the urban V2X radio channel. However, we have identified limitations of the ray tracing model in terms of multi bounce diffuse scattering. In this context, non-specular reflections are currently only taken into account for the first order (single bounce). This presented analysis has revealed that multi bounce diffuse scattering causes considerable power contributions in urban NLOS scenario. In term of computational complexity an extension of the ray tracing routine to capture these effects is a remarkable challenge and remains for future work. By analyzing the aforementioned metrics, both tools are evaluated on their accuracy and possible flaws of the underlying propagation models are identified. Due to a direct comparison of channel sounding data and ray-tracing simulations of a specific V2X scenario we provide enlightening insights which propagation effects cannot be neglected when modeling a V2X intersection scenario. Based on the results obtained from this analysis the main goal of the authors is develop a robust and flexible propagation model for the analysis of an arbitrary V2X intersection scenario. Such a model can be integrated into network simulators of other research groups and provides better approximation of a deterministic channel characterization in safety-critical urban traffic scenarios. Enabling finally the cost-efficient and flexible simulation of realistic vehicular communication scenarios might be an important driver for the worldwide development of future V2X applications.

\section*{Acknowledgment}

This work has been carried out in close collaboration between Lund University and TU Braunschweig within the COST IC 1004 framework. The authors would like to thank the IC 1004 committee for supporting and funding this short-term scientific mission that was hosted by the Institut f\"{u}r Nachrichtentechnik, Technische Universit\"{a}t Braunschweig, Braunschweig, Germany. We would also like to thank Delphi Deutschland GmbH, Bad Salzdetfurth, Germany for their contribution to the measurement campaign.

\bibliographystyle{IEEEtran}
\nocite{*}
\bibliography{STSM_Braunschweig_Journal_Version_Manuscript_arxiv.bbl}

\begin{thebibliography}{10}
\providecommand{\url}[1]{#1}
\csname url@samestyle\endcsname
\providecommand{\newblock}{\relax}
\providecommand{\bibinfo}[2]{#2}
\providecommand{\BIBentrySTDinterwordspacing}{\spaceskip=0pt\relax}
\providecommand{\BIBentryALTinterwordstretchfactor}{4}
\providecommand{\BIBentryALTinterwordspacing}{\spaceskip=\fontdimen2\font plus
\BIBentryALTinterwordstretchfactor\fontdimen3\font minus
  \fontdimen4\font\relax}
\providecommand{\BIBforeignlanguage}[2]{{%
\expandafter\ifx\csname l@#1\endcsname\relax
\typeout{** WARNING: IEEEtran.bst: No hyphenation pattern has been}%
\typeout{** loaded for the language `#1'. Using the pattern for}%
\typeout{** the default language instead.}%
\else
\language=\csname l@#1\endcsname
\fi
#2}}
\providecommand{\BIBdecl}{\relax}
\BIBdecl

\bibitem{Wiesbeck2007}
W.~Wiesbeck and S.~Knorzer, ``Characteristics of the mobile channel for high
  velocities,'' in \emph{Electromagnetics in Advanced Applications, 2007. ICEAA
  2007. International Conference on}, 2007, pp. 116--120.

\bibitem{molisch09-CommMag}
A.~F. Molisch, F.~Tufvesson, J.~Karedal, and C.~F. Mecklenbr\"{a}uker, ``A
  survey on vehicle-to-vehicle propagation channels,'' in \emph{{IEEE} Wireless
  Commun. Mag.}, vol.~16, no.~6, 2009, pp. 12--22.

\bibitem{Matolak08}
I.~Sen and D.~W. Matolak, ``Vehicle-vehicle channel models for the 5\,{GHz}
  band,'' \emph{{IEEE} Trans. Intell. Transp. Syst.}, vol.~9, no.~2, pp.
  235–--245, Jun. 2008.

\bibitem{Paier2010a}
A.~Paier, L.~Bernado, J.~Karedal, O.~Klemp, and A.~Kwoczek, ``Overview of
  vehicle-to-vehicle radio channel measurements for collision avoidance
  applications,'' in \emph{Vehicular Technology Conference (VTC 2010-Spring),
  IEEE 71st, Taipei, Taiwan}, May 2010, pp. 1--5.

\bibitem{Nuckelt11c}
J.~Nuckelt, M.~Schack, and T.~K\"{u}rner, ``Deterministic and stochastic
  channel models implemented in a physical layer simulator for {Car-to-X}
  communications,'' \emph{Advances in Radio Science}, vol.~9, pp. 165--171,
  Sept. 2011.

\bibitem{Mangel011-3}
T.~Mangel, O.~Klemp, and H.~Hartenstein, ``5.9\,{GHz} inter-vehicle
  communication at intersections: a validated non-line-of-sight path-loss and
  fading model,'' \emph{EURASIP Journal on Wireless Communications and
  Networking}, vol. 2011, no.~1, p. 182, 2011.

\bibitem{SommerWONS2011}
C.~Sommer, D.~Eckhoff, R.~German, and F.~Dressler, ``A computationally
  inexpensive empirical model of {IEEE} 802.11p radio shadowing in urban
  environments,'' in \emph{Wireless On-Demand Network Systems and Services
  (WONS), 2011 Eighth International Conference on}, 2011, pp. 84--90.

\bibitem{Hosseini2011}
S.~Hosseini~Tabatabaei, M.~Fleury, N.~Qadri, and M.~Ghanbari, ``Improving
  propagation modeling in urban environments for vehicular ad hoc networks,''
  \emph{Intelligent Transportation Systems, IEEE Transactions on}, vol.~12,
  no.~3, pp. 705--716, 2011.

\bibitem{Gaugel12}
T.~Gaugel, L.~Reichardt, J.~Mittag, T.~Zwick, and H.~Hartenstein,
  ``\BIBforeignlanguage{English}{Accurate simulation of wireless vehicular
  networks based on ray tracing and physical layer simulation},'' in
  \emph{\BIBforeignlanguage{English}{High Performance Computing in Science and
  Engineering '11}}.\hskip 1em plus 0.5em minus 0.4em\relax Springer Berlin
  Heidelberg, 2012, pp. 619--630.

\bibitem{Chelli2009}
A.~Chelli and M.~Patzold, ``The impact of fixed and moving scatterers on the
  statistics of {MIMO} vehicle-to-vehicle channels,'' in \emph{Vehicular
  Technology Conference, 2009. VTC Spring 2009. IEEE 69th}, 2009, pp. 1--6.

\bibitem{Zajic2009a}
A.~Zajic and G.~Stuber, ``Three-dimensional modeling and simulation of wideband
  mimo mobile-to-mobile channels,'' \emph{Wireless Communications, IEEE
  Transactions on}, vol.~8, no.~3, pp. 1260--1275, 2009.

\bibitem{Santos12}
R.~Santos, A.~Edwards, and V.~Rangel-Licea, \emph{Wireless Technologies in
  Vehicular Ad Hoc Networks: Present and Future Challenges}.\hskip 1em plus
  0.5em minus 0.4em\relax Hershey, PA: {IGI} Global (701 E. Chocolate Avenue,
  Hershey, Pennsylvania, 17033, USA), 2012.

\bibitem{Maurer2004}
J.~Maurer, T.~Fugen, T.~Schafer, and W.~Wiesbeck, ``A new inter-vehicle
  communications (ivc) channel model,'' in \emph{Vehicular Technology
  Conference, 2004. VTC2004-Fall. 2004 IEEE 60th}, vol.~1, 2004, pp. 9--13 Vol.
  1.

\bibitem{Joerg2013}
J.~Nuckelt, T.~Abbas, F.~Tufvesson, C.~F. Mecklenbr\"{a}uker, L.~Bernado, and
  T.~K\"{u}rner, ``{Comparison of ray tracing and channel-sounder measurements
  for vehicular communications},'' in \emph{2013 IEEE 77th Vehicular Technology
  Conference: VTC2013-Spring, Dresden, Germany}, June 2013, pp. 1--5.

\bibitem{GoogleEarth}
\BIBentryALTinterwordspacing
Google earth v7.1.1.1888 (2013). [Online]. Available:
  \url{http://www.google.com/earth/index.html [Accessed: 2013/08/15]}
\BIBentrySTDinterwordspacing

\bibitem{Schack11}
M.~Schack, J.~Nuckelt, R.~Geise, L.~Thiele, and T.~K\"{u}rner, ``Comparison of
  path loss measurements and predictions at urban crossroads for {C2C}
  communications,'' in \emph{5th European Conference on Antennas and
  Propagation (EuCAP), Rome, Italy}, April 2011.

\bibitem{Abbas2013ITST}
T.~Abbas, A.~Thiel, T.~Zemen, C.~F. Mecklenbr\"{a}uker, and F.~Tufvesson,
  ``Validation of a non-line-of-sight path-loss model for {V2V} communications
  at street intersections,'' in \emph{13th International Conference on ITS
  Telecommunications, Tampere, Finland}.\hskip 1em plus 0.5em minus 0.4em\relax
  IEEE, November 2013.

\bibitem{karedal10-05}
J.~Karedal, F.~Tufvesson, T.~Abbas, O.~Klemp, A.~Paier, L.~Bernad\'{o}, and
  A.~F. Molisch, ``Radio channel measurements at street intersections for
  vehicle-to-vehicle safety applications,'' in \emph{IEEE VTC 71st Vehicular
  Technology Conference (VTC 2010-spring), Taipei, Taiwan}, May 2010, pp. 1--5.

\bibitem{thiel10-04}
A.~Thiel, O.~Klemp, A.~Paier, L.~Bernad\'{o}, J.~Karedal, and A.~Kwoczek,
  ``In-situ vehicular antenna integration and design aspects for
  vehicle-to-vehicle communications,'' in \emph{Antennas and Propagation
  (EuCAP), 2010 Proceedings of the Fourth European Conference on, Barcelona,
  Spain}, Apr. 2010, pp. 1--5.

\bibitem{McKown91}
J.~McKown and J.~Hamilton, R.L., ``Ray tracing as a design tool for radio
  networks,'' \emph{IEEE Network}, vol.~5, no.~6, pp. 27--30, 1991.

\bibitem{Schack13}
M.~Schack, ``Integrated simulation of communication applications in vehicular
  environments,'' Ph.D. dissertation, Technische Universit\"{a}t Braunschweig,
  2013.

\bibitem{Degli-Esposti01}
V.~Degli-Esposti, ``A diffuse scattering model for urban propagation
  prediction,'' \emph{Antennas and Propagation, IEEE Transactions on}, vol.~49,
  no.~7, pp. 1111--1113, Jul 2001.

\bibitem{NuckeltEUCAP2013}
J.~Nuckelt, D.~Rose, T.~Jansen, and T.~Kurner, ``On the use of openstreetmap
  data for v2x channel modeling in urban scenarios,'' in \emph{Antennas and
  Propagation (EuCAP), 2013 7th European Conference on}, 2013, pp. 3984--3988.

\bibitem{Zheda2012}
Z.~Li, R.~Wang, and A.~Molisch, ``Shadowing in urban environments with
  microcellular or peer-to-peer links,'' in \emph{Antennas and Propagation
  (EUCAP), 2012 6th European Conference on}, 2012, pp. 44--48.

\bibitem{Molisch1999}
A.~Molisch and M.~Steinbauer, ``\BIBforeignlanguage{English}{Condensed
  parameters for characterizing wideband mobile radio channels},''
  \emph{\BIBforeignlanguage{English}{International Journal of Wireless
  Information Networks}}, vol.~6, no.~3, pp. 133--154, 1999.

\bibitem{Vaughan2003}
R.~Vaughan and J.~B. Andersen, \emph{{Channels, Propagation and Antennas for
  Mobile Communications (IEE Electromagnetic Waves Series, 50)}}.\hskip 1em
  plus 0.5em minus 0.4em\relax Institution of Engineering and Technology, Feb.
  2003.

\bibitem{molisch05}
A.~Molisch, \emph{{Wireless Communications}}.\hskip 1em plus 0.5em minus
  0.4em\relax Chichester, West Sussex, UK: IEEE Press-Wiley, 2005.

\end{thebibliography}

\end{document}